%% file: ASEGP.tex
\documentclass[11pt]{article}

\usepackage{natbib}
\usepackage{graphicx}
\usepackage{amssymb}
\usepackage{amsmath}
\usepackage{bm}
\usepackage{amsthm}
\usepackage{amsfonts}
\usepackage{hyperref}
\usepackage{xr}
\usepackage{dsfont}
\usepackage{authblk}
\usepackage{times}
\usepackage[margin=0.8in]{geometry}

\newcommand{\E}{\mathbb{E}}
\newcommand{\Var}{\mathrm{Var}}

\newcommand\blfootnote[1]{%
\begingroup
\renewcommand\thefootnote{}\begin{NoHyper}\footnote{#1}%
\addtocounter{footnote}{-1}\end{NoHyper}%
\endgroup
}

\title{\vspace{-0.7in}Adjusting for Spatial Effects in Genomic Prediction}
\author[1]{Xiaojun Mao}
\author[2]{Somak Dutta}
\author[3]{Raymond K. W. Wong}
\author[2]{Dan Nettleton}
\affil[1]{School of Data Science, Fudan University, Shanghai 200433, China.}
\affil[2]{Department of Statistics, Iowa State University, Ames, IA 50011, USA}
\affil[3]{Department of Statistics, Texas A\&M University, College Station, TX 77843, USA}

\date{}

\begin{document}
\maketitle
\vspace{-0.7in}\begin{abstract}
    This paper investigates the problem of adjusting for spatial effects in genomic prediction. Despite being seldomly considered in genomic prediction, spatial effects often affect phenotypic measurements of plants. We consider a Gaussian random field model with an additive covariance structure that incorporates genotype effects, spatial effects and subpopulation effects. An empirical study shows the existence of spatial effects and heterogeneity across different subpopulation families, while simulations illustrate the improvement in selecting genotypically superior plants by adjusting for spatial effects in genomic prediction.
\end{abstract}
\blfootnote{E-mail addresses: maoxj@fudan.edu.cn, somakd@iastate.edu, raywong@tamu.edu, dnett@iastate.edu.}
\section{Introduction}\label{sec:intro}
In plant breeding, predicting the genetic value of plant genotypes plays an important role in determining which genotypes to include in subsequent generations. Recently, several powerful statistical methods have been developed that use high-dimensional single-nucleotide polymorphism (SNP) genotypes for genomic prediction. Most of the methods based on mixed linear models (MLM) are quite flexible due to the consideration of fixed and random effects. For instance, population structure (discussed in \citealt{Pritchard-Stephens-Rosenberg00}) is often accounted for by modeling the fixed effects of principal components (PCs) derived from the SNPs \citep{Price-Patterson-Plenge06,Reich-Price-Patterson08,McVean09}. For unified MLM approaches \citep{Yu-Pressoir-Briggs06}, SNP data are used to determine a kinship matrix that is assumed to be proportional to the variance of a vector of random effects that accounts for dependencies due to relatedness among individuals. For a more computationally efficient compressed MLM (CMLM) approach \citep{Zhang-Ersoz-Lai10}, data from many individuals are compressed into a smaller number of groups, and the interindividual kinship matrix is replaced by a lower-dimensional matrix that characterizes correlations among group random effects induced by genetic similarities among groups.

Aside from correlations due to relatedness among individuals or groups, phenotypes measured on plants grown in fields can be spatially correlated \citep{stroup2002power}. Such correlation can arise because plants growing near each other may share a common microenvironment that differs from the microenvironment experienced by plants in other parts of the field. This microenvironmental variation can induce phenotypic similarity among neighboring plants. When such spatial effects exist but are unaccounted for in the analysis, decisions about which plant genotypes are expected to perform best with regard to one or more phenotypic traits can be adversely affected. With the adjustment of these effects, some high-throughput phenotyping technologies \citep{Cabrera-Bosquet-Crossa-Zitzewitz12,Masuka-Araus-Das12, White-Andrade-Sanchez-Gore12} can be applied to increase plant yields.

Several works \citep{stroup1991nearest,stroup1994removing,Gilmour-Cullis-Verbyla97,Crossa-Burgueno-Cornelius06,Lado-Matus-Rodriguez13,Bernal-Vasquez-Mohring-Schmidt14,Selle-Steinsland-Hickey19} have considered spatial effects in linear mixed-effects models. \citet{Gilmour-Cullis-Verbyla97} proposed a separable autoregressive (AR$\times$AR) model. As suggested by \citet{Bernal-Vasquez-Mohring-Schmidt14}, fitting a row and column model (RC) (i.e., a model with an effect for each row and for each column in a field experiment layout) can account for a substantial portion of phenotypic heterogeneity that may be due to spatial effects. \citet{Lado-Matus-Rodriguez13} compared RC models with approaches that attempt to adjust for spatial effects by using the difference between a plot's response value and the average response of its neighboring plots as a covariate. Such a method, referred to by \citet{Lado-Matus-Rodriguez13} as ``moving-means as a covariate'' (MVNG), was found to best fit the data and lead to the most accurate phenotypic predictions. \citet{Selle-Steinsland-Hickey19} considered a Bayesian framework to model the spatial mixed-effects. In addition, several smoothing methods \citep{Verbyla-Cullis-Kenward99,Durban-Hackett-McNicol03,Rodriguez-Alvarez-Boer-Eeuwijk18} are developed to model the spatial trends more explicitly. In particular, \citet{Rodriguez-Alvarez-Boer-Eeuwijk18} used a smooth bivariate surface (SpATS) to model the random spatial effects which captures both large-scale and small-scale dependences. In this paper, we propose an alternative modeling strategy that has some conceptual advantages because it integrates information from high-dimensional markers, spatial locations and subpopulation structure via Gaussian kernels and provides competing or better performance than several existing approaches for spatial adjustments in genomic prediction.

We study two datasets. One is a maize dataset involving a nested association mapping (NAM) panel consisting of 4660 recombinant inbred lines (RILs) derived from crosses between a reference inbred line B73 and 25 other founder inbreds. More information about the NAM panel is available in \citet{Yu-Holland-McMullen08} and at \url{http://www.panzea.org}. The RILs derived by crossing B73 to any one of the 25 other founders from a subpopulation of RILs. Thus, the 4660 RILs we consider can be partitioned into 25 subpopulations. Even after conditioning on SNP genotypes carried by each RIL, phenotypic responses from RILs within a subpopulation are expected to be more strongly correlated than responses from RILs in different subpopulations. This within-subpopulation correlation is expected due to shared genetic material as well as characteristics of the experimental design described in Section~\ref{sec:data}. The second dataset is a wheat dataset which consists of genotype and phenotype data on 384 advanced lines from two different breeding programs. The data are provided in \citet{Lado-Matus-Rodriguez13}.

The goal of this paper is to predict the genetic value of each maize RIL or each wheat line from a huge number of SNP marker genotypes, while accounting for the genetic and spatial dependence among phenotypic measurements. We focus on a Gaussian random field (GRF) model with an additive covariance matrix structure that incorporates genotype effects, spatial effects and subpopulation effects. For genotype effects, we adopt a Gaussian kernel \citep{Morota-Koyama-Rosa13,Ober-Erbe-Long11} to capture general relationships between genotypes and phenotypes. We compare our spatially adjusted genomic predictions with genomic predictions generated by a design-based incomplete block (IB) linear mixed-effects model and existing methods CMLM \citep{Zhang-Ersoz-Lai10}, RC \citep{Bernal-Vasquez-Mohring-Schmidt14}, MVNG \citep{Lado-Matus-Rodriguez13} and SpATS \citep{Rodriguez-Alvarez-Boer-Eeuwijk18}. In a simulation study presented in Section \ref{sec:sim}, we apply the proposed GRF method to help identify the best plant genotypes.

The rest of the paper is organized as follows. Real data are described in Section \ref{sec:data}. The proposed GRF is constructed in Section \ref{sec:methods}. Within Section \ref{sec:methods}, we also discuss kernels and corresponding parameter estimation methods. Numerical performances of the proposed method for genomic predictions and for choosing the best plant genotypes are illustrated in an empirical study in Section \ref{sec:real} and a simulation study in Section \ref{sec:sim}, respectively. The paper concludes with a discussion in Section \ref{sec:discussion}. Some supplementary materials are provided in Section \ref{appsec:preprocess}.

\begin{figure}[htp]
\centering
\includegraphics[scale=0.7]{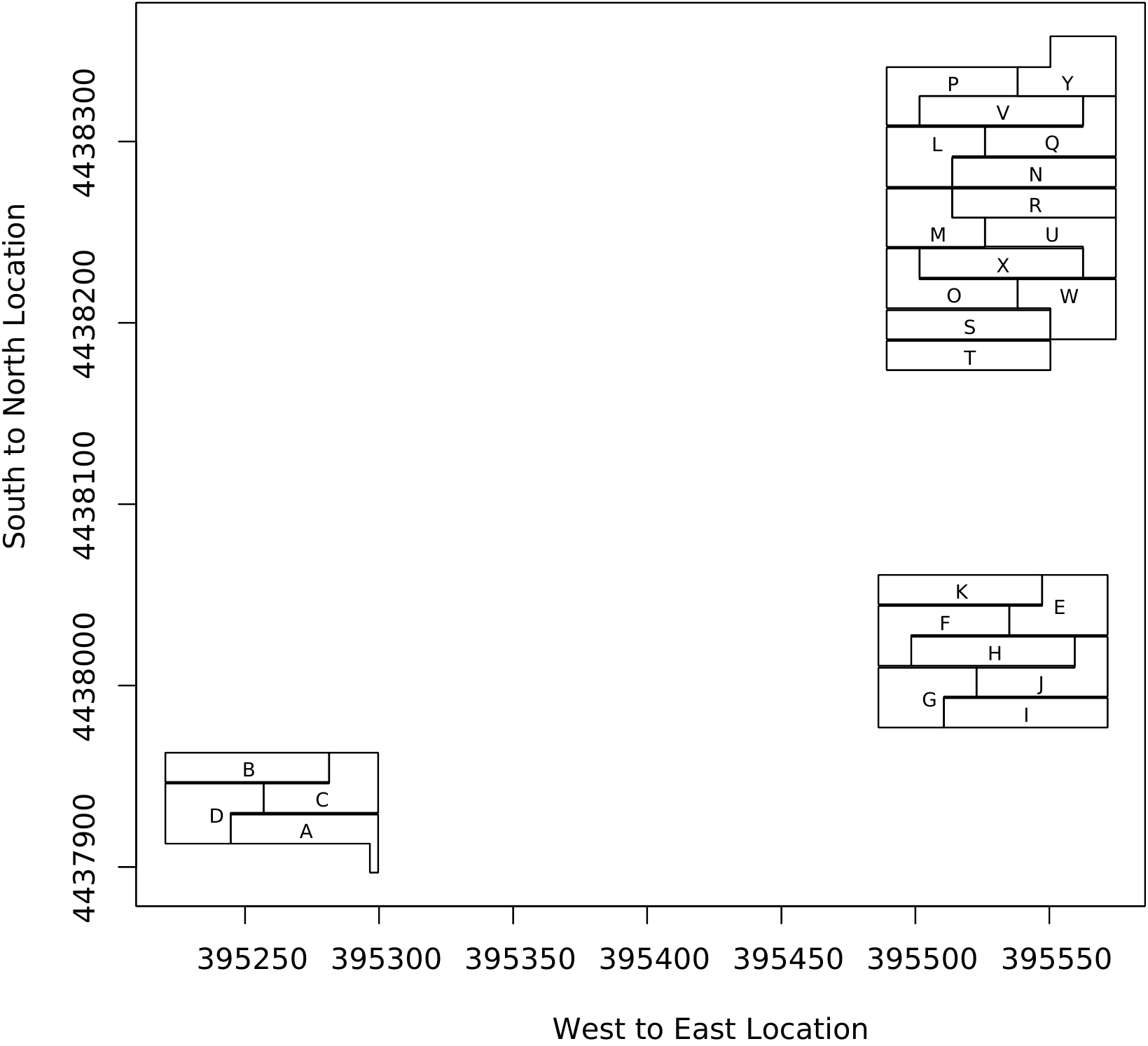}
\caption{Geographic locations of 25 subpopulations labeled A through Y.}
\label{fig:subinfo}
\end{figure}

\section{Data}\label{sec:data}
Throughout this paper, we used \textsf{Data1} to refer to a maize NAM RIL dataset comprised of 4660 RILs genotyped at $687,869$ SNP markers.  The phenotypic value for each RIL is a measurement of the carbon dioxide ($\text{CO}_2$) emitted from plant material incorporated in a soil sample. Scientific interest centers on identifying RILs whose genetic constitution makes them relatively low emitters of $\text{CO}_2$.

The 4660 RILs in \textsf{Data1} can be partitioned into 25 subpopulations, each produced from a biparental cross of inbred line B73 to one of the 25 NAM founder inbred lines. A side-by-side box plot to compare the carbon dioxide emissions among these 25 subpopulations is available in Figure \ref{appfig:carbonsub} of the supplemental document. Due to the large number of RILs and limited field plot availability, the experimental design is unreplicated with a single plot for each RIL distributed across three nearby agricultural fields, with no two plots separated by more than 2.5 miles.  RILs from any given subpopulation were randomized to plots within a single subpopulation-specific region (as depicted in Figure \ref{fig:subinfo}) to facilitate mapping of quantitative trait loci separately for each subpopulation. In our combined analysis of data from all RILs, we expect correlations among the phenotypic values for RILs within each subpopulation due to both region and subpopulation effects, as well as spatially correlated plot effects, which induce correlations among phenotypic values within any field regardless of subpopulation membership.   

Our second dataset (henceforth labeled \textsf{Data2}) is the 2011 wheat dataset presented in \citet{Lado-Matus-Rodriguez13}. This dataset contains results for  384 wheat lines genotyped at 102324 biallelic markers and phenotyped for grain yield (GY), thousand kernel weight (TKW), the number of kernels per spike (NKS), and days to heading (DH) under two levels of water supply: mild water stress (MWS) and fully irrigated (FI).  For both MWS and FI, the 384 wheat lines were planted in an alpha-lattice design with $20$ incomplete blocks of size 20 and two complete replications.  Within each replications, 382 genotypes were planted on one plot each, while 2 of the 384 genotypes were planted on 9 plots each to cover all $20 \times 20$ plots. \citet{Lado-Matus-Rodriguez13} also analyzed data collected in 2012 from two separate locations, but the 2012 data contain measurements of only the grain yield phenotype.  We restrict our analysis to the 2011 data only to simplify our presentation. The same dataset has been analyzed by \citet{Rodriguez-Alvarez-Boer-Eeuwijk18} and \citet{Selle-Steinsland-Hickey19}.    

\section{Methods}\label{sec:methods}
\subsection{Models}\label{sec:models}
We are given a training dataset $\{y_{i}, \bm{x}_i, b_i, \bm{s}_i\}_{i=1}^n$, where $y_i\in\mathbb{R}$ represents a phenotype measurement, $\bm{x}_i\in\mathcal{X}$ is the corresponding $p$-dimensional vector of binary marker genotypes, $b_i\in\mathcal{B}$ is the corresponding subpopulation family index of the observation and $\bm{s}_i\in\mathcal{S}$ is the corresponding spatial location of the observation. Here, $\mathcal{X}$, $\mathcal{B}$ and $\mathcal{S}$ represent the sets of possible values of binary marker genotype vectors, subpopulation family indices and spatial locations, respectively.

We propose a Gaussian random field (GRF) approach that carefully models (i) genotype effects, (ii) subpopulation effects and (iii) spatial effects.
More specifically, for $i=1,\ldots,n$, suppose
\begin{equation}
y_{i} = \mathbb{Z}\left(\bm{t}_i\right) + \epsilon_{i}\label{eqn:basic_model},
\end{equation}
where $\bm{t}_i= (\bm{x}^\intercal_i, b_i, \bm{s}^\intercal_i)^\intercal$, $\mathbb{Z}(\bm{t}_{i})$ is an observation at $\bm{t}_{i}$ of a GRF $\mathbb{Z}$ defined over index domain $\mathcal{T}=\mathcal{X}\times \mathcal{B}\times\mathcal{S}$, and $\epsilon_{i}$ is a mean-zero Gaussian random variable independent of $\mathbb{Z}$. Further, we  let $\bm{\epsilon}=(\epsilon_{1}, \ldots, \epsilon_{n})^\intercal$ and assume $\Var(\bm{\epsilon})=\sigma_{\epsilon}^2\bm{I}_{n\times n}$ with $\bm{I}_{n\times n}$ being the identity matrix of size $n$. We assume a constant mean function for $\mathbb{Z}$, i.e., $\E(\mathbb{Z}(\bm{t}))=\mu$ for any $\bm{t}\in\mathcal{T}$. The power of this model lies in the flexible modeling of the covariance structure of $\mathbb{Z}$.

We consider an additive model for the covariance function
that accounts for the three major effects. Specifically, for any $\bm{t}_{i}=(\bm{x}_{i}^\intercal, b_{i}, \bm{s}_{i}^\intercal)^\intercal,
\bm{t}_{k}=(\bm{x}_{k}^\intercal, b_{k}, \bm{s}_{k}^\intercal)^\intercal \in
\mathcal{T}=\mathcal{X}\times\mathcal{B}\times\mathcal{S}$, we assume
\[
\text{Cov}[\mathbb{Z}(\bm{t}_i),\mathbb{Z}(\bm{t}_k)]=C\left(\bm{t}_{i}, \bm{t}_{k}\right) = \sigma_{g}^{2}C_{g}\left(\bm{x}_{i}, \bm{x}_{k}\right)+\sigma_{b}^{2}C_{b}\left(b_{i},b_{k}\right)+\sigma_{s}^{2} C_{s}\left(\bm{s}_{i},\bm{s}_{k}\right),
\]
where $\sigma_{g}^{2}$, $\sigma_{b}^{2}$ and $\sigma_{s}^{2}$ are variance components and $C_{g}:\mathcal{X}^2\to\mathbb{R}$, $C_{b}:\mathcal{B}^2\to\mathbb{R}$ and $C_{s}:\mathcal{S}^2\to\mathbb{R}$ are unit-diagonal kernel functions that quantify the corresponding dependence structures arising from similarity among observations with respect to genetic markers, subpopulations and spatial locations, respectively. Equivalently, we assume that the GRF $\mathbb{Z}$ can be decomposed into $\mathbb{Z}(\bm{t}_i)=\mu+\mathbb{Z}_{g}(\bm{x}_i)+\mathbb{Z}_{b}(b_i)+\mathbb{Z}_{s}(\bm{s}_i)$, where $\mathbb{Z}_{g}$, $\mathbb{Z}_{b}$, $\mathbb{Z}_{s}$ are mean-zero Gaussian random fields with covariance structures determined by $\sigma_{g}^{2}{C}_{g}$, $\sigma_{b}^{2}{C}_{b}$ and $\sigma_{s}^{2}{C}_{s}$, respectively. We quantify the strength of spatial effects relative to the effects associated with marker genotypes by the variance component ratio $\gamma=\sigma_{s}^{2}/\sigma_{g}^{2}$.

\subsection{Marker Kernel \texorpdfstring{$C_{g}$}{Cg}}\label{sec:markerk}
Following \citet{Morota-Gianola14} and references therein,
we choose the Gaussian kernel
\[
C_{g}\left(\bm{x}_{i}, \bm{x}_{k}\right) = \exp\left( -\frac{\|\bm{x}_{i}-\bm{x}_{k}\|^2}{\tau} \right),\text{ for any } \bm{x}_{i},\bm{x}_{k}\in\mathcal{X}
\]
where $\|\cdot\|$ represents the Euclidean norm and $\tau$ is a parameter greater than zero.

Compared with other common kernels, the Gaussian kernel has been empirically shown to give robust and strong predictive performance. In \citet{Ober-Erbe-Long11}, the more general Mat\'ern kernel is studied, but the Gaussian kernel performed best among the Mat\'ern family based on their simulation study. Since the marker genotypes take discrete values, there is a temptation to choose a kernel on discrete index space. In \citet{Morota-Koyama-Rosa13}, a discretized Gaussian kernel, referred to as a diffusion kernel, was applied to dairy and wheat data for predicting phenotypes using marker information. However, the predictive power of such a kernel was similar to the Gaussian kernel.

Current high-throughput genotyping technology can provide genotype calls for hundreds of thousands of SNPs. Since most SNPs are unassociated with phenotype or conditionally unassociated with phenotype given other SNPs, $C_{g}(\bm{x}_{i}, \bm{x}_{k})$ does not necessarily provide a good representation of correlation between the $i$-th and $k$-th lines when all SNPs are included in the vector of marker genotypes. To reduce computation time and improve genomic prediction, we use $\mathrm{FarmCPU}$ \citep{Liu-Huang-Fan16} to select important SNPs for inclusion in $\bm{x}_{i}$ rather than using the entire ensemble of SNPs. The details of our SNP selection procedure are discussed in Section \ref{appsec:preprocess} of the supplementary material.

\subsection{Subpopulation Kernel \texorpdfstring{$C_{b}$}{Cb}}\label{sec:subpopk}
The subpopulation GRF $\mathbb{Z}_{b}$ is motivated by genetic heterogeneity across different subpopulations and genetic similarity within subpopulations that may not be fully captured by SNP genotypes. We consider $C_{b}(b_{i},b_{k})=\mathds{1}(b_{i}=b_{k})$ for any $b_{i},b_{k}\in\mathcal{B}$, where $\mathds{1}(\cdot)$ is the indicator function. This covariance structure is equivalent to that induced by a model with independent, constant-variance subpopulation random effects.

\subsection{Spatial Kernel $C_{s}$}\label{sec:spatialk}
In an agricultural field trial, plots are typically embedded in a regular rectangular array with say $m_1$ rows and $m_2$ columns. To adjust for spatial effects that may exist in such trials, the class of spatial autoregressions on regular rectangular lattice has been quite popular following the works of \cite{Besag-Green93,Besag-Green-Higdon95,Besag-Higdon99,Dutta-Mondal15} and \cite{Mondal-Dutta-etal20}. In this work we focus on the class of stationary autoregressions with modification described as follows. Consider a bigger array with $m_1' = m_1 + 4$ rows and $m_2' = m_2 + 4$ columns obtained by adding two virtual plots \citep{Besag-Higdon99} to each boundary to reduce the boundary effects. Suppose that for any positive integer $k,$ $\bm{W}_k$ denotes the $k\times k$ matrix with
\[
W_k(1,1) = W_k(k,k) = 1,~W_k(i,i) = 2 (1 < i <k),~W_k(i,i+1) = W_k(i+1,i) = -\mathds{1}(1\leq i < k),
\]
and $W_k(i,j) = 0$ otherwise.
Here, $ W_k(i,j)$ represents the $(i,j)$-th entry of $\bm{W}_k$.
Next define $\bm{N}_{01} = \bm{I}_{m_2^{\prime}}\otimes \bm{W}_{m_1^{\prime}}$,  $ \bm{N}_{10} = \bm{W}_{m_2^{\prime}}\otimes \bm{I}_{m_1^{\prime}}$,
and
\[\bm{W} = \beta_{00}\bm{I} + \beta_{01}\bm{N}_{01} + \beta_{10}\bm{N}_{10},\]
where $\beta_{00}$, $\beta_{01}$ and $\beta_{10}$ are positive parameters with the identifiability constraint $\beta_{00} + 2(\beta_{01} + \beta_{10}) = 1.$ Let $\bm{D}$ be the diagonal matrix consisting of the diagonal entries of $\bm{W}^{-1}.$ Next, for any plot $s_i$ suppose $\bm{h}_i = \bm{h}(s_i)$ denotes the incidence vector of length $m_1'm_2'$. That is,  the $j$-th entry of $\bm{h}_i$ is 1 if and only if $s_i$ corresponds to the $j$th plot in the $m_1'\times m_2'$ array where the plots in the array are enumerated in a column major format; the rest of the entries of $\bm{h}_i$ are zeros.

Finally, the spatial covariance of $\left(\mathbb{Z}_s(s_1),\ldots,\mathbb{Z}_s(s_n)\right)^\intercal$ is given by
\[
\sigma_s^2C_s(s_i,s_j) = \sigma^2_s\bm{h}_i^\intercal\bm{D}^{-1/2}\bm{W}^{-1}\bm{D}^{-1/2}\bm{h}_j,
\]
where $\sigma^2_s$ is the spatial variance component introduced in Section~\ref{sec:methods}.
The advantage of this spatial kernel is that it makes the marginal variances at observed plots constant, while keeping the pairwise correlations the same as those that would be obtained from the stationary autoregression covariance matrix $\bm{W}^{-1}$. However, note that the weights on the neighbors are no longer $\beta_{01}$ and $\beta_{10}$ but are approximately proportional to them at the interior plots because the variances at the interior plots are approximately constant \citep{Besag-Kooperberg95}.

The anisotropy parameters $\beta_{01}$ and $\beta_{10}$ play an important role because these parameters are related to the field geometry. In fact, \citet{McCullagh-Clifford06} found substantial empirical evidence that the non-anthropogenic variability in field trials can be explained by an isotropic spatial process with correlation decaying approximately logarithmically with distance. This would imply, for example, that for square plots the values of $\beta_{01}$ and $\beta_{10}$ should be approximately equal. On the other hand, if the plots are rectangular and the spacing between the plots is negligible compared with the plot sizes, the ratio of the $\beta_{01}$ and $\beta_{10}$ should be close to the aspect ratio of the plots. Also if the design is single column replication (see the El Bat\'an trial in \citet{Besag-Higdon99}), then $\beta_{10}$ is \emph{zero}. In practice the estimates of $\beta_{01}$ and $\beta_{10}$ are automatically adjusted to the plot geometry and the interplot spacing. For \textsf{Data1}, in our context, the spatial layout in the maize experiment mimics a single--column replicate design because the distance between two east--west neighboring maize plants is much larger than the distance between two north--south neighbors. Thus, we apriori expect the estimate of $\beta_{10} \approx 0$. This expectation is corroborated by the ML estimates in Section \ref{sec:maize data}, where we see that the MLE of $\beta_{10}$ occurs at the boundary. On the other hand, the plots in the \textsf{Data2} are rectangular, and the interplot spacings are not very large. Thus, we expect the MLE of $\beta_{10}$ and $\beta_{01}$ to be somewhere between $0$ and $0.5$ and this is corroborated by the estimates in Section \ref{sec:wheat data}.

The parameter $\beta_{00}$, on the other hand, controls the strength of the neighboring correlations and the range of the correlation. Interestingly, the boundary value of $\beta_{00} = 0$ gives rise to an intrinsic autoregression process and is the focus of \citet{Besag-Green-Higdon95}, \citet{Besag-Higdon99} and \citet{Dutta-Mondal15} in the context of fertility adjustments in agricultural variety trials. In particular, the foundational work of \citet{McCullagh-Clifford06} and empirical evidence from \citet{Besag-Green-Higdon95}, \citet{Besag-Higdon99} and \citet{Dutta-Mondal15,Dutta-Mondal16} advocate the use of the intrinsic model for spatial adjustments in agricultural trials. Consequently, to build a proper covariance model and to avoid boundary issues in maximum likelihood estimation, we fix the parameter $\beta_{00}$ at a small value. To that end, following the suggestion in \citet{Besag-Kooperberg95}, we numerically compute the neighboring correlations for various values of $\beta_{00}$ (with $\beta_{01} = \beta_{10}$ = $(1-\beta_{00})/2)$. We observe that the theoretical neighboring correlation changes by 11.26\% when $\beta_{00}$ changes from $0.01$ to $0.001$ and only by 1.69\% when the $\beta_{00}$ changes from $0.001$ to $0.0001$. A similar conclusion is obtained where $\beta_{01}$ is held fixed at other values including $0$ and $0.5$. Because a change of 1.69\% in neighboring correlation is practically negligible, we choose to fix the value of $\beta_{00}$ at $0.001$. Our analyses (see supplementary materials Table \ref{apptab:lambda}) also shows that the prediction accuracies and ranking do not change appreciably by changing $\beta_{00}$ from $0.001$ to $0.0001$.

We end this section with references to other commonly used spatial kernels in such prediction problems. Rather than using the autoregression models to fit the spatial effects, several works \citep{Crossa-Burgueno-Cornelius06,Lado-Matus-Rodriguez13,Bernal-Vasquez-Mohring-Schmidt14} considered them as random effects in simple mixed linear models. In the context of agricultural field trials, \citet{Gleeson-Cullis87}, \citet{Cullis-Gleeson91}, \citet{Zimmerman-Harville91}, \citet{Gleeson-Cullis87}, \citet{Gilmour-Cullis-Verbyla97} and \citet{Cullis-Gogel-Verbyla98} developed  elaborate  model  for adjusting for systematic and non-systematic trends.  Their spatial adjustment models, despite drawing  criticism  for  their  heavy  dependence  on  the  coordinate  system  by  \citet{McCullagh-Clifford06}, have been quite effective in practice for spatial adjustments and may be potential alternatives for spatial adjustments in genomic prediction.

\subsection{Estimation}
For the training dataset $\{y_{i}, \bm{x}_i, b_i, \bm{s}_i\}_{i=1}^n$, let $\bm{C}$ be the $n\times n$ matrix with element $i,j$ equal to $C(\bm{t}_{i},\bm{t}_{j})$ for all $i,j=1,\ldots,n$. Define $\bm{C}_{g}$, $\bm{C}_{b}$ and $\bm{C}_{s}$ analogously. Then, the covariance matrix of the vector of $n$ phenotypic response values $\bm{y}=(y_1,\ldots,y_n)^{\intercal}$ can be written as
\[
\bm{\Sigma} = \bm{C}+\sigma_{\epsilon}^{2}\bm{I}_{n\times n}=\sigma_{g}^{2}\bm{C}_{g}+\sigma_{b}^{2}\bm{C}_{b}+\sigma_{s}^{2} \bm{C}_{s}+\sigma_{\epsilon}^{2}\bm{I}_{n\times n}.
\]
The variance--covariance matrix $\bm{\Sigma}$ is a function of the parameters $\sigma_{g}$, $\tau$, $\sigma_{b}$, $\sigma_{s}$ and $\sigma_{\epsilon}$. We maximize the log-likelihood to estimate these five parameters simultaneously. 

It is straightforward to show that, for any given value of $\bm{\Sigma}$, the likelihood is maximized over $\mu$ at $\widehat{\mu}=\bm{1}^{T}\bm{\Sigma}^{-1}\bm{y}/\bm{1}^{T}\bm{\Sigma}^{-1}\bm{1}$. Thus, the corresponding profile log-likelihood function is 
\[
\ell\left(\sigma_{g},\tau,\sigma_{b},\sigma_{s},\sigma_{\epsilon}\right)=-\frac{1}{2}\log|\bm{\Sigma}|-\frac{\left(\bm{y}-\widehat{\mu}\bm{1}\right)^{T}\bm{\Sigma}^{-1}\left(\bm{y}-\widehat{\mu}\bm{1}\right)}{2}.
\]
Finding maximizers of this profile log-likelihood function yields maximum likelihood estimates (MLEs) $\widehat{\sigma}_{g},\widehat{\tau},\widehat{\sigma}_{b},\widehat{\sigma}_{s}$, and $\widehat{\sigma}_{\epsilon}$. Let $\widehat{\bm{C}}_{g}$ and $\widehat{\bm{\Sigma}}$ be the estimates of the covariance structures $\bm{C}_{g}$ and $\bm{\Sigma}$ obtained by replacing the unknown parameters with their MLEs.

Considering the joint distribution of $\bm{y}$, $\bm{Z}_{g}=(\mathbb{Z}_{g}(\bm{x}_{1}),\ldots,\mathbb{Z}_{g}(\bm{x}_{n}))^{\intercal}$, $\bm{Z}_{b}=(\mathbb{Z}_{b}(b_{1}),\ldots,\mathbb{Z}_{b}(b_{n}))^{\intercal}$ and $\bm{Z}_{s}=(\mathbb{Z}_{s}(\bm{s}_{1}),\ldots,\mathbb{Z}_{s}(\bm{s}_{n}))^{\intercal}$, we have
\[
{\small 
    \begin{bmatrix}
    \bm{y} \\
    \bm{Z}_{g} \\
    \bm{Z}_{b} \\
    \bm{Z}_{s}
    \end{bmatrix}
    \sim
    \mathcal{N}\left(
    \begin{bmatrix}
    \mu\bm{1} \\
    \bm{0} \\
    \bm{0} \\
    \bm{0}
    \end{bmatrix},
    \begin{bmatrix}
    \bm{\Sigma} &
    \sigma_{g}^{2}\bm{C}_{g} & \sigma_{b}^{2}\bm{C}_{b} & \sigma_{s}^{2}\bm{C}_{s} \\
    \sigma_{g}^{2}\bm{C}_{g} & \sigma_{g}^{2}\bm{C}_{g} & \bm{0} & \bm{0} \\
    \sigma_{b}^{2}\bm{C}_{b} & \bm{0} & \sigma_{b}^{2}\bm{C}_{b} & \bm{0} \\
    \sigma_{s}^{2}\bm{C}_{s} & \bm{0} & \bm{0} &  \sigma_{s}^{2}\bm{C}_{s}
    \end{bmatrix}\right).
}
\]

Based on our MLEs, we can estimate the conditional mean and conditional variance of $\bm{Z}_{g}$, $\bm{Z}_{b}$ and $\bm{Z}_{s}$ given $\bm{y}$, by
\begin{equation}\label{eqn:expect}
\widehat{\E}\left(\left.\begin{bmatrix}
\bm{Z}_{g} \\
\bm{Z}_{b}\\
\bm{Z}_{s}
\end{bmatrix}\right|\bm{y} \right)=
\begin{bmatrix}
\bm{0} \\
\bm{0} \\
\bm{0}
\end{bmatrix}+
\begin{bmatrix}
\widehat{\sigma}_{g}^{2}\widehat{\bm{C}}_{g}\\
\widehat{\sigma}_{b}^{2}\bm{C}_{b}\\
\widehat{\sigma}_{s}^{2}\bm{C}_{s}
\end{bmatrix}
\widehat{\bm{\Sigma}}^{-1}
(\bm{y}-\widehat{\mu}\bm{1})
\end{equation}
and
\begin{equation}\label{eqn:var}
\widehat{\Var}\left(\left.\begin{bmatrix}
\bm{Z}_{g} \\
\bm{Z}_{b} \\
\bm{Z}_{s}
\end{bmatrix}\right|\bm{y}\right)=
\begin{bmatrix}
\widehat{\sigma}_{g}^{2}\widehat{\bm{C}}_{g} & 0 & 0 \\
0 & \widehat{\sigma}_{b}^{2}\bm{C}_{b} & 0 \\
0 & 0 & \widehat{\sigma}_{s}^{2}\bm{C}_{s}
\end{bmatrix}
-
\begin{bmatrix}
\widehat{\sigma}_{g}^{2}\widehat{\bm{C}}_{g} \\
\widehat{\sigma}_{b}^{2}\bm{C}_{b} \\
\widehat{\sigma}_{s}^{2}\bm{C}_{s}
\end{bmatrix}
\widehat{\bm{\Sigma}}^{-1}
\begin{bmatrix}
\widehat{\sigma}_{g}^{2}\widehat{\bm{C}}_{g} & \widehat{\sigma}_{b}^{2}\bm{C}_{b} &
\widehat{\sigma}_{s}^{2}\bm{C}_{s}
\end{bmatrix}.
\end{equation}

\section{Empirical Study}\label{sec:real}
\subsection{Existing Methods}\label{sec:exmethod}
For the purpose of benchmarking, we compared our method with methods based on the Compressed Mixed Linear Model (CMLM) \citep{Zhang-Ersoz-Lai10} implemented in the \textsf{R} package GAPIT \citep{Lipka-Tian-Wang12}, the Row and Column Model (RC) \citep{Bernal-Vasquez-Mohring-Schmidt14}, the linear regression with moving means as covariable model (MVNG) \citep{Lado-Matus-Rodriguez13} and Spatial Analysis of field Trials with Splines (SpATS) implemented in the \textsf{R} package SpATS \citep{Rodriguez-Alvarez-Boer-Eeuwijk18}. These competing methods are described in the following. All the methods are implemented in \textsf{R} language.

\subsubsection*{Compressed Mixed Linear Model}
Let $\bm{M}$ be a matrix whose columns correspond to the first few principal components (usually 3 or 5 by default) computed from the binary genotype matrix to represent population structure. The compressed mixed linear model is
\[
\bm{y}=\mu\bm{1}+\bm{M}\bm{\beta}+\bm{Z}\bm{\bar{u}}+\bm{e},
\]
where $\bm{\bar{u}}_{r\times 1}\sim\mathcal{N}(\bm{0},\sigma_{\bar{u}}^{2}\bm{\bar{K}}_{r\times r})$ represents an unknown vector of random additive genetic effects and $\bm{e}\sim\mathcal{N}(\bm{0},\sigma_{e}^{2}\bm{I})$ is the unobserved vector of errors. The random effects in $\bm{\bar{u}}_{r\times 1}$ are intended to represent the effects of multiple background quantitative trait loci (QTL) on the phenotypic response values. Note that $\bm{\bar{u}}_{r\times 1}$ is of dimension $r\times 1$ rather than $n\times 1$ as in the MLM because that $\bm{\bar{u}}_{r\times 1}$ represents different groups $t=1,\ldots,r$ clustered according to a full kinship matrix $\bm{K}_{n\times n}$ rather than individuals/lines. Meanwhile, the matrix $\bm{\bar{K}}_{r\times r}$ is the corresponding kinship matrix that accounts for varying degrees of genetic similarity among groups rather than among individuals/lines. We adopt the formula for the full kinship matrix suggested by \citet{VanRaden08}:
\begin{equation}
\bm{K}_{n\times n}=\frac{\widetilde{\bm{X}}^{(g)}\widetilde{\bm{X}}^{(g) \intercal}}{\sum_{i}2p_i(1-p_i)},
\label{eqn:fullkinship}
\end{equation}
where $\widetilde{\bm{X}}^{(g)}$ contains allele calls centered so that each row sums to zero and $p_i$ is the frequency of the minor allele at locus $i$. As for the group kinship matrix $\bm{\bar{K}}_{r\times r}=(\bar{K}_{st})$ where $s,t=1$ to $r$, each of the entry $\bar{K}_{st}$ is defined as the average of a set of $\{K_{hj}\}$ where $h$ belongs to group $s$ and $j$ belongs to group $t$. For the maize dataset \textsf{Data1}, the Bayesian information criterion \citep{Zhang-Ersoz-Lai10} selects no principal components in the matrix $\bm{M}$. For the wheat dataset \textsf{Data2}, we considered one, three, five and ten principal components for $\bm{M}$. We found no important difference, and thus, we adopted the default setting with the first three principal components in $\bm{M}$.

\subsubsection*{ Incomplete Block Model}
Motivated by the alpha-lattice experimental design underlying the wheat dataset,
we also consider an incomplete block (IB) model defined as follows.
Using the same principal component matrix $\bm{M}$ in CMLM,
the IB model assumes
\[
\bm{y}=\mu\bm{1}+\bm{M}\bm{\beta}+\bm{Z}_{\bm{u}_{g}}\bm{u}_{g}+\bm{Z}_{\bm{u}_{\text{rep}}}\bm{u}_{\text{rep}}+\bm{Z}_{\bm{u}_{\text{bl(rep)}}}\bm{u}_{\text{bl(rep)}}+\bm{e},
\]
where $\bm{u}_{g}\sim\mathcal{N}(\bm{0},\sigma_{g}^{2}\bm{K})$,
$\bm{u}_{\text{rep}}\sim\mathcal{N}(\bm{0},\sigma_{\text{rep}}^{2}\bm{I})$,
$\bm{u}_{\text{bl(rep)}}\sim\mathcal{N}(\bm{0},\sigma_{\text{bl(rep)}}^{2}\bm{I})$ and 
$\bm{e}\sim\mathcal{N}(\bm{0},\sigma_{e}^{2}\bm{I})$
represent independent, unknown vectors of additive genetic effects,
replication effects, incomplete block effects and errors, respectively.
Here, $\bm{K}$ is the full kinship matrix defined in \eqref{eqn:fullkinship}.
We applied this model to the wheat dataset \textsf{Data2} with the first three principal components in $\bm{M}$.
Because the experimental design that gave rise to the maize dataset involves no replication or blocking, the IB model is not applicable for \textsf{Data1}.

\subsubsection*{RC and MVNG}
For the Row and Column Model (RC) and the linear regression with moving means as covariate model (MVNG), we propose two steps for the prediction as suggested by \citet{Lado-Matus-Rodriguez13}. The idea is that we first adjust for spatial effects in the observed phenotypic response values, and then we provide genomic predictions by using the \textsf{R} package $\mathrm{rrBLUP}$ applied to the spatially adjusted phenotypic response values. Two different kernels, RR and GAUSS \citep{Endelman11}, are considered for the genomic predictions.

In the first step, the RC model assumes that
\[
y_{ijk}=\mu+\text{row}_i+\text{col}_j+\text{sub}_k+e_{ijk},
\]
where $\text{row}_i$ (row effect), $\text{col}_j$ (column effect) and $\text{sub}_k$ (subpopulation effect) are considered as independent random effects with mean-zero normal distributions that have variances specific to the effect type (i.e., one variance for row effects, one for column effects and one for subpopulation effects). For the adjustment, we have $\widehat{y}_{ijk}=y_{ijk}-\widehat{\text{row}}_i-\widehat{\text{col}}_j$ where $\widehat{\text{row}}_i=\widehat{\E}(\text{row}_i|\bm{y})$ and $\widehat{\text{col}}_j=\widehat{\E}(\text{col}_j|\bm{y})$ are the corresponding empirical Best Linear Unbiased Predictors (eBLUPs) of $\text{row}_i$ and $\text{col}_j$ effects.

For MVNG, we adopt the same idea in \citet{Lado-Matus-Rodriguez13}, namely, we fit the model
\[
y_{i}=\mu+\beta x_i+e_{i},
\]
where $x_i=y_i-\frac{1}{6}\sum_{k=1}^{6}y_i^{(k)}$ with $y_i^{(k)}$, $k=1,\ldots,6$, the phenotypic response values for the spatial neighbors (one up, one down, two left, and two right) of the $i$-th observation (See Figure 1 in \citet{Lado-Matus-Rodriguez13} for details). For \textsf{Data2}, as suggested by \citet{Lado-Matus-Rodriguez13}, left--right corresponds to spatial neighbors within each row and up--down corresponds to spatial neighbors within each column. For \textsf{Data1}, based on the observation that east--west neighbors are much farther apart than north--south neighbors, we adopt north--south as left--right and east--west as up--down in this MVNG method. The spatially adjusted values for $i$-th observation is given by $\widehat{y}_{i}=y_{i}-\widehat{\beta}x_i$.

In the second step, the genomic prediction is performed under the model
\[
\widehat{\bm{y}}=\mu\bm{1}+\bm{Z}\bm{u}+\bm{e},
\]
where $\bm{u}\sim\mathcal{N}(\bm{0},\sigma_{u}^{2}\bm{K})$ represents an unknown vector of random additive genetic effects and $\bm{e}\sim\mathcal{N}(\bm{0},\sigma_{e}^{2}\bm{I})$ is the unobserved vector of residuals. For kernel RR, $\bm{K}=\bm{X}\bm{X}^{\intercal}$, where $\bm{X}$ is the original genotype matrix without scaling and centering. For kernel GAUSS, $\bm{K}=\bm{C}_{g}$, the parameter $\tau$ is estimated by residual maximum likelihood (REML).

\subsubsection*{SpATS}
For the Spatial Analysis of field Trials with Splines,
we consider a spatial model defined as follows.
Using the same principal component matrix $\bm{M}$ in CMLM,
the IB model assumes
\[
\bm{y}=f(\bm{v},\bm{u})+\bm{Z}_{\bm{c}_{g}}\bm{c}_{g}+\bm{Z}_{\bm{c}_{\text{r}}}\bm{c}_{\text{r}}+\bm{Z}_{\bm{c}_{\text{c}}}\bm{c}_{\text{c}}+\bm{e},
\]
where $\bm{c}_{g}\sim\mathcal{N}(\bm{0},\sigma_{g}^{2}\bm{I})$,
$\bm{c}_{\text{r}}\sim\mathcal{N}(\bm{0},\sigma_{\text{r}}^{2}\bm{I})$,
$\bm{c}_{\text{c}}\sim\mathcal{N}(\bm{0},\sigma_{\text{c}}^{2}\bm{I})$ and 
$\bm{e}\sim\mathcal{N}(\bm{0},\sigma_{e}^{2}\bm{I})$
represent independent, unknown vectors of additive genetic effects,
row effects, column effects and errors, respectively.
Here, $ f(\bm{v},\bm{u})$ is a smooth bivariate surface modeled by P-splines \citep{Rodriguez-Alvarez-Boer-Eeuwijk18}. We applied this model to the wheat dataset \textsf{Data2}. In its current implementation, SpATS cannot predict new observations whose row or column is out of the existing row or column grid. Because the fields and subpopulations in \textsf{Data1} are arranged in irregular grids, we were not able to apply the SpATS approach for \textsf{Data1}.

\subsection{\textsf{Data1} Prediction}\label{sec:maize data}
As described in Section \ref{sec:data}, the maize dataset (\textsf{Data1}) can be naturally divided into three fields or into 25 subpopulations (see Figure \ref{fig:subinfo}). In this section, we provide evidence of both spatial effects and subpopulation effects in each field and evidence of spatial effects in each subpopulation. To provide such evidence, we fit three reduced versions of the full GRF model defined in Sections \ref{sec:markerk}--\ref{sec:spatialk}. For the dataset in each field, we fit both $\text{GRF}_{-Z_{b}}$ and $\text{GRF}_{-Z_{s}}$, where the corresponding covariances are $\bm{\Sigma}_{-Z_{b}}=\sigma_{g_{-Z_{b}}}^{2}\bm{C}_{g}+\sigma_{s_{-Z_{b}}}^{2} \bm{C}_{s}+\sigma_{\epsilon_{-Z_{b}}}^{2}\bm{I}$ and $\bm{\Sigma}_{-Z_{s}}=\sigma_{g_{-Z_{s}}}^{2}\bm{C}_{g}+\sigma_{b_{-Z_{s}}}^{2} \bm{C}_{b}+\sigma_{\epsilon_{-Z_{s}}}^{2}\bm{I}$, respectively; i.e., we ignore subpopulation effects in $\text{GRF}_{-Z_{b}}$ and spatial effects in $\text{GRF}_{-Z_{s}}$. For any dataset consisting of a single subpopulation, we drop subpopulation effects and fit $\text{GRF}_{-Z_{bs}}$ instead of $\text{GRF}_{-Z_{s}}$, where $\bm{\Sigma}_{-Z_{bs}}=\sigma_{g_{-Z_{bs}}}^{2}\bm{C}_{g}+\sigma_{\epsilon_{-Z_{bs}}}^{2}\bm{I}$.

In the following, we report the performance of CMLM, RC(RR, GAUSS), MVNG(RR, GAUSS), $\text{GRF}_{-Z_{b}}$, $\text{GRF}_{-Z_{s}}$, $\text{GRF}_{-Z_{bs}}$ and the full GRF based on analysis of $1000$ independent random partitions of the data in each subpopulation into training ($80\%$) and test ($20\%$) sets. When performing analysis at the field level, we combine the training sets from all subpopulations in a field to form one training set and likewise pool the corresponding subpopulation-specific test sets to form a field-specific test set.

To evaluate the performance of different methods, we consider the accuracy defined as the correlation between predicted response values and observed phenotypic response values in the test set. In Table \ref{tab:fieldresult}, we report the accuracies for each field, along with estimates of $\widehat{\gamma}=\widehat{\sigma}_{s}^2/\widehat{\sigma}_{g}^2$ based on the whole dataset (without splitting). Due to space limitation, the detailed results for each subpopulation are relegated to  Table \ref{apptab:fieldsubresult} of the supplementary material. The magnitude of $\widehat{\gamma}$ indicates the estimated strength of spatial effects relative to genotypic variation. 	As we can see in Table \ref{tab:fieldresult} (and also Table \ref{apptab:fieldsubresult}), the GAUSS kernel is inferior to the RR kernel in both RC and MVNG results. Thus, we present only RC(RR) and MVNG(RR) results in subsequent figures.

\begin{figure}[!h]
    \centering
    \includegraphics[scale=0.4]{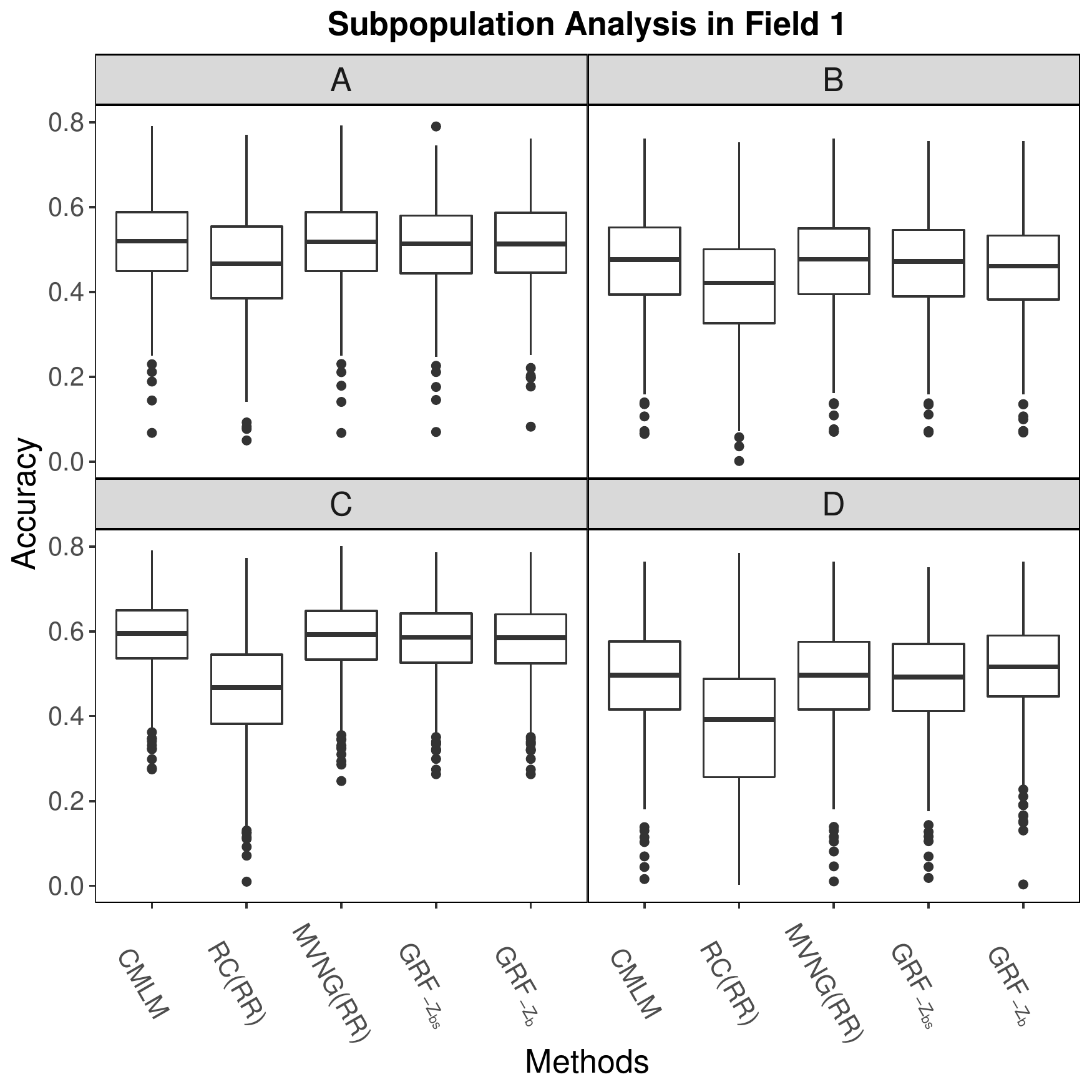}
    \includegraphics[scale=0.4]{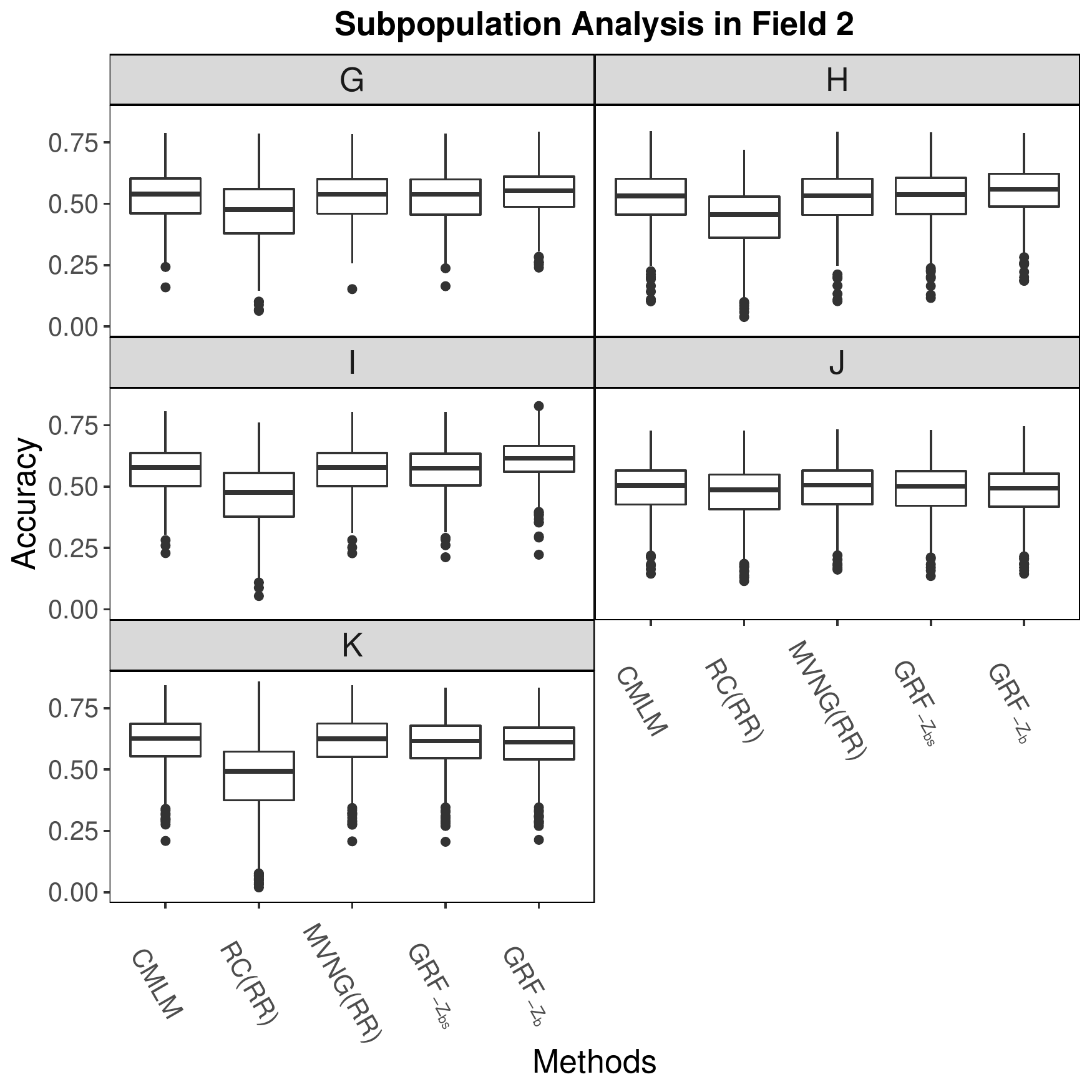}
    \includegraphics[scale=0.4]{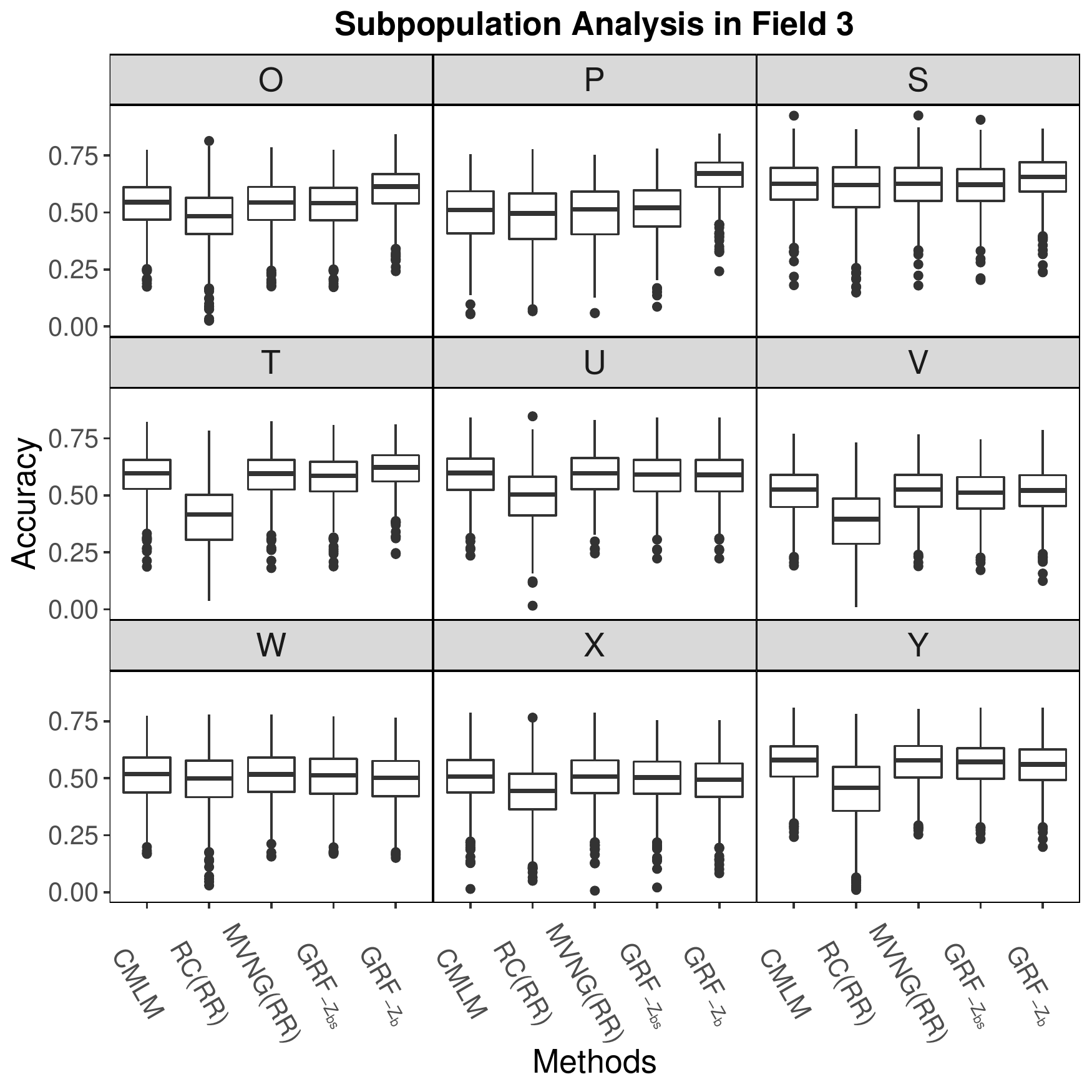}
    \includegraphics[scale=0.4]{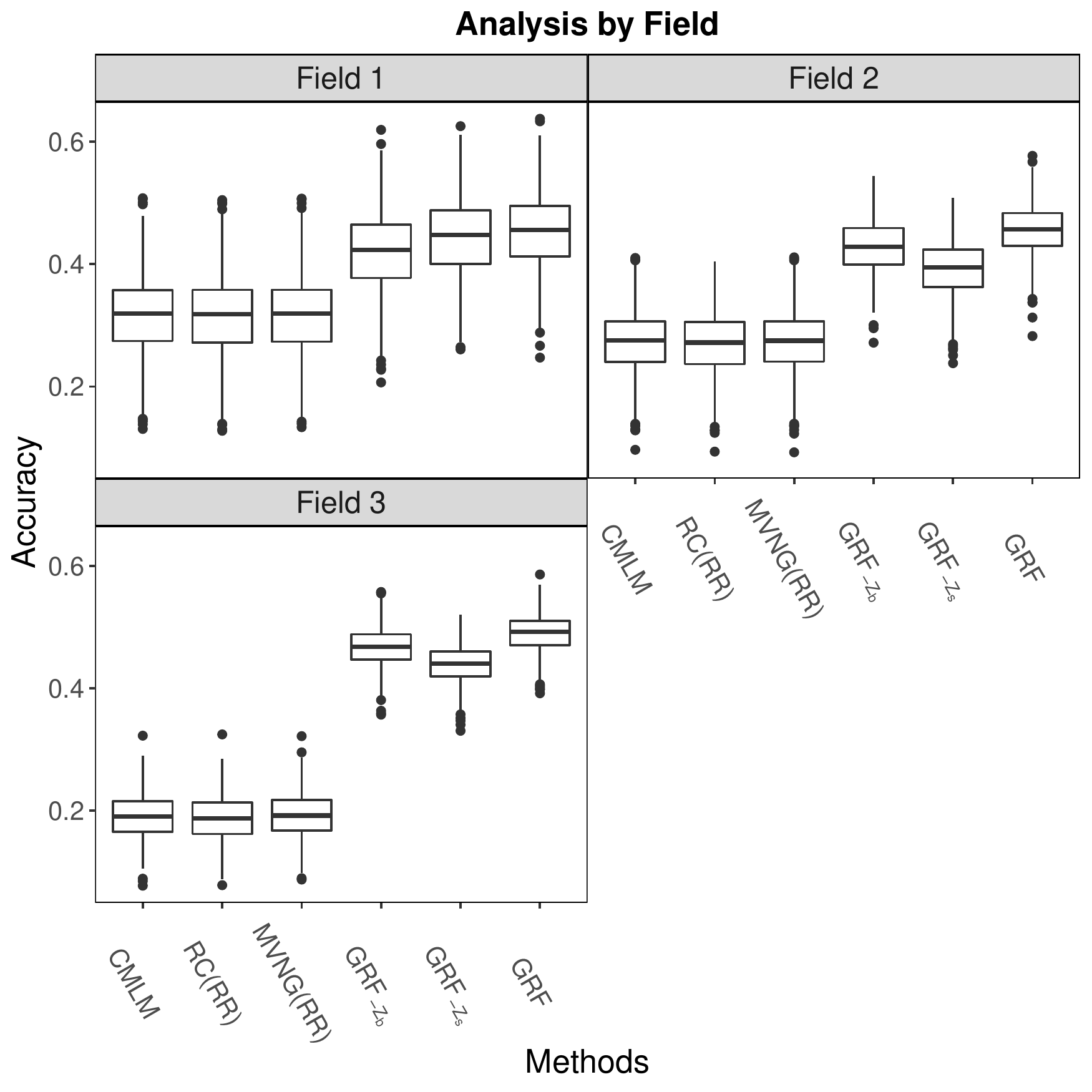}
    \caption{(1--3): Comparison of CMLM, RC(RR), MVNG(RR) and two proposed methods $\text{GRF}_{-Z_{bs}}$ and $\text{GRF}_{-Z_{b}}$ for each subpopulation (only $9$ of $14$ are shown in Field $3$ to improve clarity). (4): Comparison of CMLM, RC(RR), MVNG(RR) and three proposed methods $\text{GRF}_{-Z_{b}}$, $\text{GRF}_{-Z_{s}}$ and GRF for each field.}
    \label{fig:gwas}
\end{figure}

For each subpopulation, Figure \ref{fig:gwas} (1--3) shows the comparison of CMLM, RC(RR), MVNG(RR) and the two proposed methods $\text{GRF}_{-Z_{bs}}$ and $\text{GRF}_{-Z_{b}}$.  RC(RR) exhibits noticeably lower accuracy than the other methods for most of the subpopulations.  The accuracy distributions appear similar for the other methods in most subpopulations.  The proposed method $\text{GRF}_{-Z_{b}}$ stands out as the best performing method for a few subpopulations, especially in Field 3 (see, e.g., subpopulations O and P).  From Table~\ref{apptab:fieldsubresult}, we see that $\text{GRF}_{-Z_{b}}$ has the highest average accuracy for more than half of the 25 subpopulations.  When the accuracy of $\text{GRF}_{-Z_{b}}$ is close to or lower than accuracies of existing methods, the estimated strength of spatial effects $\widehat{\gamma}$ is close to $0$. For the subpopulations with strong spatial effects, it is reasonable that the predictions can be improved relative to CMLM (which ignores spatial effects) by incorporating the spatial kernel $\bm{C}_{s}$. Because there is little evidence of horizontal spatial correlation, RC(RR) and MVNG(RR) are based on misspecified spatial models which lead to lower accuracy.

For the subpopulations with weak or no spatial effects, accuracy of predictions may be degraded by inclusion of $\bm{C}_{s}$ in the model. Comparing CMLM and $\text{GRF}_{-Z_{bs}}$ (the methods that ignore spatial effects), we can see that $\text{GRF}_{-Z_{bs}}$ has slightly lower average accuracies for many subpopulations.	A possible explanation is that
CMLM makes greater use of the SNP information.
While SNP information enters the marker kernel of $\text{GRF}_{-Z_{bs}}$ via simple Euclidean distances,
CMLM utilizes this information in both fixed effects and random effects. Specifically, CMLM allows for fixed effects of the PCs of SNPs and adopts the corresponding kinship matrix as the variance--covariance structure for random effects. This may also be the reason that the GAUSS kernel is inferior or similar to the RR kernel in both RC and MVNG methods.

For the field-level analysis, we are able to use the full GRF that includes genotype, subpopulation and spatial effects. Figure \ref{fig:gwas} (4) and Table \ref{tab:fieldresult} show that the full GRF has the highest average accuracy across all methods for every field. These results illustrate that the full GRF can effectively account for heterogeneity across genotype, subpopulation and spatial location effects at the field scale to enhance prediction accuracy.

\begin{table}[t]
    \caption{Average accuracies and their standard deviations (in parentheses) for \textsf{Data1} by five existing methods and three proposed methods ($\text{GRF}_{-Z_{b}}$, $\text{GRF}_{-Z_{s}}$ and GRF) for each field based on $1000$ independent random partitions of the data into training ($80\%$) and test ($20\%$) sets.}
    \label{tab:fieldresult}
    {\footnotesize
    \begin{center}
    \begin{tabular}{|c|c|c|c|c|c|}
        \cline{1-6}
        & \multicolumn{4}{|c|}{Method} & \\
        \cline{1-6}
        \hline
        Field & \multicolumn{2}{|c|}{RC} & \multicolumn{2}{|c|}{MVNG} & $\widehat{\gamma}=\widehat{\sigma}_{s}^{2}/\widehat{\sigma}_{g}^{2}$ \\ 
        \hline
        &  RR & GAUSS & RR & GAUSS &  \\ 
        \hline
        1 & 0.3173 (0.0633) & 0.3144 (0.0628) & 0.3159 (0.0638) & 0.3131 (0.0633) & 0.0646 \\ 
        2 & 0.2727 (0.051) & 0.2729 (0.0511) & 0.2697 (0.0501) & 0.2698 (0.0501) & 0.3041 \\ 
        3 & 0.1913 (0.0362) & 0.1904 (0.0363) & 0.1883 (0.0361) & 0.1873 (0.0363) & 1.0087 \\ 
        \hline
        & CMLM & $\text{GRF}_{-Z_{b}}$ & $\text{GRF}_{-Z_{s}}$ & GRF &  \\ 
        \hline
        1 & 0.3173 (0.0632) & 0.4199 (0.0641) & 0.4428 (0.064) & $\bm{0.4520}$ (0.0629) & \\ 
        2 & 0.2727 (0.0501) & 0.4289 (0.0422) & 0.3920 (0.0464) & $\bm{0.4558}$ (0.0410) & \\ 
        3 & 0.1930 (0.0361) & 0.4672 (0.0301) & 0.4395 (0.0305) & $\bm{0.4904}$ (0.0296) & \\ 
        \hline
    \end{tabular}\\The highest average accuracy across methods for each field is shown in bold.
    \end{center}}
    
\end{table}

\subsection{\textsf{Data2} Prediction}\label{sec:wheat data}
For the wheat dataset \textsf{Data2}, there is no subpopulation information. Thus, we do not need the component $\bm{Z}_b$ in the full GRF, and the corresponding subpopulation covariance structure $\bm{C}_b$ is ignorable. In the following, we report the performance of CMLM, IB, RC(RR, GAUSS), MVNG(RR, GAUSS), SpATS, $\text{GRF}_{-Z_{b}}$ and $\text{GRF}_{-Z_{bs}}$ based on $1000$ independent training--test partitions for the eight phenotypes in the wheat dataset \textsf{Data2}. Due to space limitation, the detailed results are relegated to Tables \ref{apptab:wheatpred1} and \ref{apptab:wheatpred2} of the supplementary material. In addition to the prediction results, the corresponding parameter estimates are reported in Table \ref{apptab:wheatspatial} in the supplementary material. For each partition, we split the dataset into training ($86\%$) and test ($14\%$) sets. To avoid overestimating the prediction accuracy for new genotypes, we split so that all observations for any genotype are contained entirely within the training set or entirely within the test set.  In contrast, completely random partitioning often distributes observations for a given genotype to both training and test sets.  Because test set predictions for such genotypes can directly utilize information about the genotype from the training set, measurements of prediction accuracy from completely random partitioning tend to be much higher than should be expected when predicting the performance of a new genotype whose phenotypic values are not part of the training data.  For this reason, our results on prediction accuracy are not comparable to the results reported in Table 3 of \citet{Lado-Matus-Rodriguez13}.  However, we do include the methods considered by \citet{Lado-Matus-Rodriguez13} in our study, with some adjustments to improve methods when possible.  In particular, \citet{Lado-Matus-Rodriguez13} presented results for an inferior-performing version of our IB approach that involved using genomic prediction techniques on the residuals from the fit of the IB model without genomic information. Results labeled IB in this paper refer to our implementation of the IB model described in Section \ref{sec:real}. All the methods are implemented in \textsf{R} language. Our \textsf{R} code is included in the online supplementary material. They can also be found at\\
\url{https://github.com/mxjki/Adjusting_for_Spatial_Effects_in_Genomic_Prediction}.

\begin{figure}[!h]
    \centering
    \includegraphics[scale=0.55]{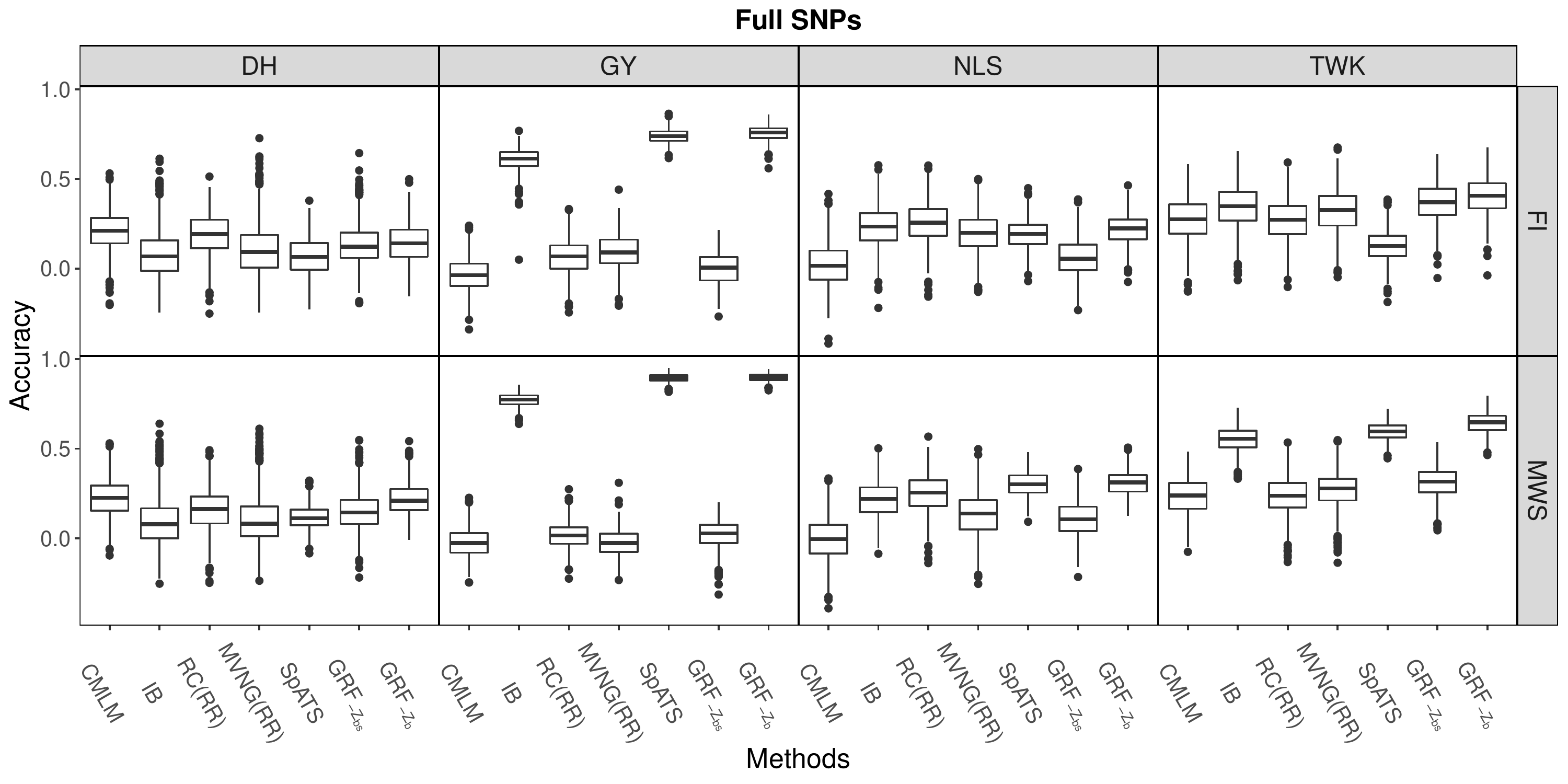}
    \includegraphics[scale=0.55]{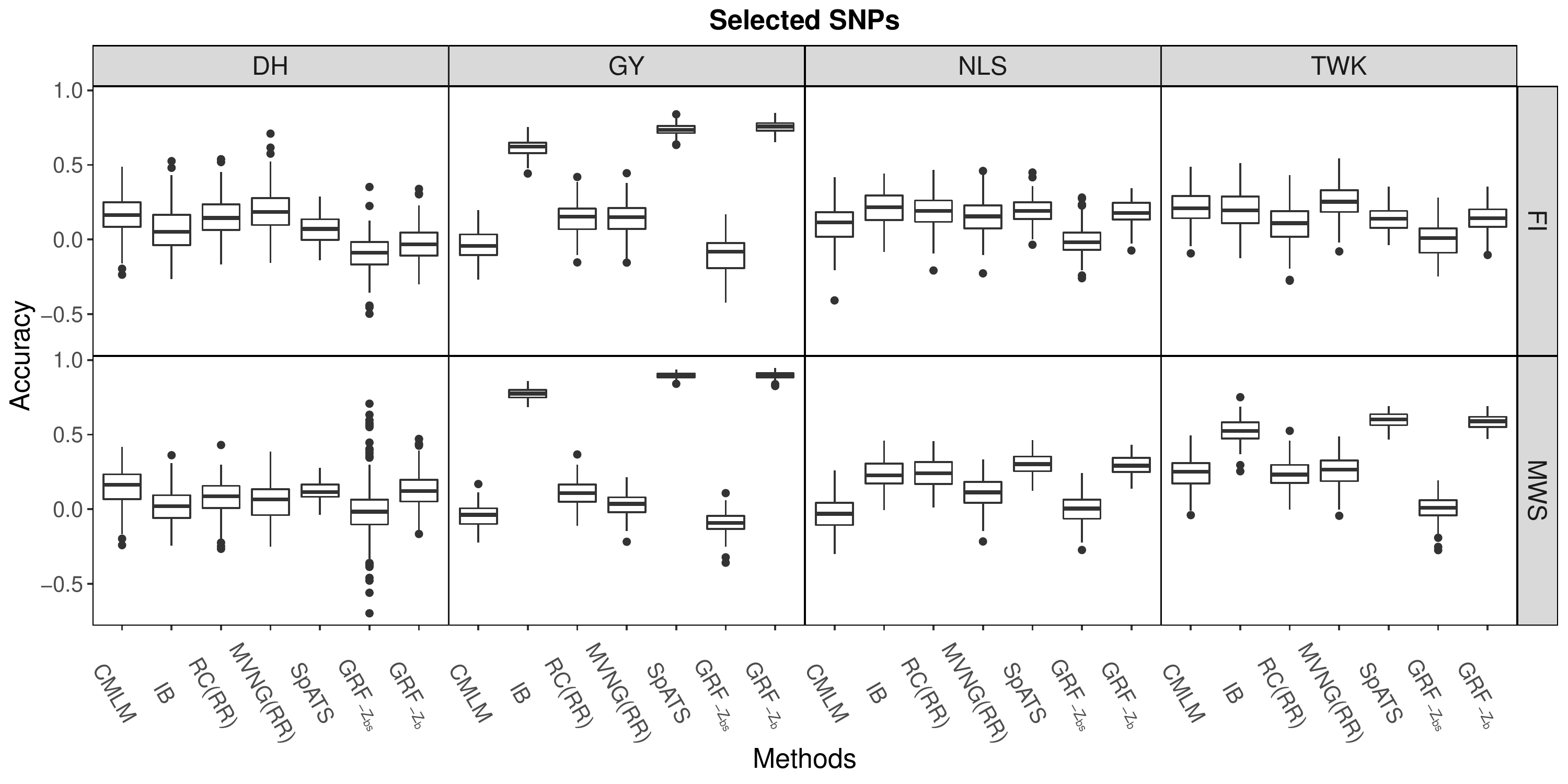}
    \caption{Comparisons of CMLM, IB, RC(RR), MVNG(RR), SpATS and two proposed methods $\text{GRF}_{-Z_{bs}}$ and $\text{GRF}_{-Z_{b}}$ with full and selected SNPs.}
    \label{fig:gwaswheat}
\end{figure}

Figure \ref{fig:gwaswheat} shows the comparison of CMLM, IB, RC(RR), MVNG(RR), SpATS, $\text{GRF}_{-Z_{b}}$ and $\text{GRF}_{-Z_{bs}}$ in terms of accuracy. To allow for a clearer visual depiction of results, RC and MVNG results based on GAUSS kernels appear only in Tables \ref{apptab:wheatpred1} and \ref{apptab:wheatpred2} in the supplemental document.  Table~\ref{apptab:wheatpred1} shows that $\text{GRF}_{-Z_{b}}$ has the highest average accuracy among all methods for five of the eight environment $\times$ phenotype combinations and higher average accuracy than the most similar competing method (SpATS) for all eight environment $\times$ phenotype combinations when using the full set of SNPs.  Spatial effects are strongest for the phenotype grain yield (GY) in Santa Rosa under two levels of water supply, mild water stress (MWS) and fully irrigated (FI), with estimated relative strength of spatial effects $\widehat{\gamma}$ is $2.4231$ and $4.5171$, respectively (see Table~\ref{apptab:wheatspatial}). Figures in \citet{Rodriguez-Alvarez-Boer-Eeuwijk18} and \citet{Selle-Steinsland-Hickey19} provide a detailed description of the strong spatial effects.  When spatial effects are strongest, Figure~ \ref{fig:gwaswheat} shows that SpATs and $\text{GRF}_{-Z_{b}}$ clearly outperform other approaches, regardless of whether full or selected SNPs are used.  When spatial effects are strong, the accuracy difference between $\text{GRF}_{-Z_{b}}$ and $\text{GRF}_{-Z_{bs}}$ is much larger than for other phenotypes. Figure \ref{fig:gwaswheat} (and also Tables \ref{apptab:wheatpred1}, \ref{apptab:wheatpred2} and \ref{apptab:wheatspatial}) indicates that although the general results are not strongly sensitive to selection of SNPs prior to model fitting and analysis, the average accuracy of $\text{GRF}_{-Z_{b}}$ tends to decrease following SNP selection.

\section{Simulation Study}\label{sec:sim}
This section reports results from simulation experiments designed to evaluate numerical performance of genomic predictions after adjusting for spatial effects.

\subsection{\textsf{Data1} Ranking}
From the maize dataset \textsf{Data1}, we fit the full GRF model to obtain parameter estimates $\widehat{\mu}, \widehat{\sigma}_{g},\widehat{\tau},\widehat{\sigma}_{b},\widehat{\sigma}_{s}$, and $\widehat{\sigma}_{\epsilon}$. These estimates provide $\widehat{\bm{C}}_{g}$ and $\widehat{\bm{\Sigma}}$, which determine the estimated mean and variance of the conditional multivariate normal distribution for $\bm{Z}_{g}$, $\bm{Z}_{b}$ and $\bm{Z}_{s}$ according to equations \eqref{eqn:expect} and \eqref{eqn:var}.
Given these estimated parameters,
let $\widetilde{\bm{Z}}_{g}$, $\widetilde{\bm{Z}}_{b}$ and $\widetilde{\bm{Z}}_{s}$ be generated simultaneously from
a multivariate normal distribution
where the mean and variance are specified in (\ref{eqn:expect}) and (\ref{eqn:var}).
And let $\tilde{\bm{e}}$ be generated from $\mathcal{N}(\bm{0},\widehat{\sigma}_{\epsilon}^{2}\bm{I})$.
To allow different strengths of spatial effects, we simulate the response vector $\tilde{\bm{y}}=\widehat{\mu}\bm{1}+\widetilde{\bm{Z}}_{g}+\widetilde{\bm{Z}}_{b}+c\widetilde{\bm{Z}}_{s}+\tilde{\bm{e}}$, where $c\in\{1,2,3,4\}$ controls the strength of spatial effects. Given a simulated dataset, we fit the full GRF, $\text{GRF}_{-Z_{s}}$ and $\text{GRF}_{-Z_{b}}$ to predict $\tilde{\bm{y}}$.
We repeat this simulation and fitting process $1000$ times.

In addition to prediction accuracy, we
also compare the ability to rank plant genotypes.
We compare the true rank-order $\bm{r}^{(o)}$ of the elements of $\widehat{\mu}\bm{1}+\widetilde{\bm{Z}}_{g}+\widetilde{\bm{Z}}_{b}$, with the rank-orders $\bm{r}^{(\text{GRF})}$, $\bm{r}^{(\text{GRF}_{-Z_{s}})}$ and $\bm{r}^{(\text{GRF}_{-Z_{b}})}$ of the predictions by computing the Spearman's rank-order correlations 
$\rho_s(\bm{r}^{(o)},\bm{r}^{(\text{GRF})})$, $\rho_s(\bm{r}^{(o)},\bm{r}^{(\text{GRF}_{-Z_{s}})})$, and $\rho_s(\bm{r}^{(o)},\bm{r}^{(\text{GRF}_{-Z_{b}})})$ for each simulation replication.

Table \ref{tab:rankorder} reports both the prediction accuracies and Spearman's rank-order correlations. These two measurements are highly correlated. The full GRF is much better than $\text{GRF}_{-Z_{s}}$ and $\text{GRF}_{-Z_{b}}$ in terms of prediction accuracies and the similarities of rank-orders with the true rank-order $\bm{r}^{(o)}$. Because spatial effects and subpopulation effects for \textsf{Data1} in each field are strong enough ($\widehat{\gamma}=0.0646$, $\widehat{\sigma}_{b}=0.3581$; $\widehat{\gamma}=0.3041$, $\widehat{\sigma}_{b}=0.3526$ and $\widehat{\gamma}=1.0087$, $\widehat{\sigma}_{b}=0.3939,$ respectively, for the three fields.) With spatial strength held constant, prediction performance in Table \ref{tab:rankorder} improves across fields in accordance with the number of observations per field, likely due to the improvement of estimation with more data.

\begin{table}[t]
    \caption{Average prediction accuracies and Spearman's rank-order correlations ($\rho_s$) based on 1000 simulations for \textsf{Data1} by the full GRF, $\text{GRF}_{-Z_{s}}$ and $\text{GRF}_{-Z_{b}}$ for different spatial strengths.}
    \label{tab:rankorder} 
    {\centering
    \small
    \begin{tabular}{|c|c|c|c|c|c|c|c|}
        \cline{1-8}
        \hline
        Field & Strength & \multicolumn{3}{|c|}{Accuracies} & \multicolumn{3}{|c|}{$\rho_s$}\\
        \hline
        & & GRF & $\text{GRF}_{-Z_{s}}$ & $\text{GRF}_{-Z_{b}}$ & GRF & $\text{GRF}_{-Z_{s}}$ & $\text{GRF}_{-Z_{b}}$ \\ 
        \hline
        1 & 1 & $\bm{0.8249}$ & 0.8200 & 0.5041 & $\bm{0.8076}$ & 0.8036 & 0.4711 \\ 
        & 2 & $\bm{0.8069}$ & 0.7853 & 0.4795 & $\bm{0.7887}$ & 0.7676 & 0.4459 \\ 
        & 3 & $\bm{0.7860}$ & 0.7343 & 0.4672 & $\bm{0.7659}$ & 0.7169 & 0.4332 \\ 
        & 4 & $\bm{0.7632}$ & 0.6738 & 0.4571 & $\bm{0.7421}$ & 0.6595 & 0.4229 \\
        \hline
        2 & 1 & $\bm{0.8395}$ & 0.8317 & 0.5221 & $\bm{0.8276}$ & 0.8196 & 0.5008 \\ 
        & 2 & $\bm{0.8129}$ & 0.7706 & 0.5070 & $\bm{0.7995}$ & 0.7563 & 0.4858 \\ 
        & 3 & $\bm{0.7847}$ & 0.6900 & 0.4952 & $\bm{0.7699}$ & 0.6762 & 0.4740 \\ 
        & 4 & $\bm{0.7554}$ & 0.6067 & 0.4836 & $\bm{0.7395}$ & 0.5956 & 0.4627 \\ 
        \hline
        3 & 1 & $\bm{0.9135}$ & 0.9085 & 0.2835 & $\bm{0.8986}$ & 0.8934 & 0.2760 \\ 
        & 2 & $\bm{0.8818}$ & 0.8505 & 0.2693 & $\bm{0.8637}$ & 0.8310 & 0.2621 \\ 
        & 3 & $\bm{0.8486}$ & 0.7738 & 0.2601 & $\bm{0.8286}$ & 0.7512 & 0.2531 \\ 
        & 4 & $\bm{0.8164}$ & 0.6940 & 0.2523 & $\bm{0.7952}$ & 0.6693 & 0.2453 \\ 
        \hline 
    \end{tabular}}\\
    The highest average accuracy and highest average rank-order correlation across methods for each combination of field and spatial strength are shown in bold.
\end{table}

For each simulated data set, we predict the top $l$ inbred lines are by ranking our predictions of $\mathbb{Z}_{g}+\mathbb{Z}_{b}$, for $l\in\{1,\ldots,n\}$.
We use $T_{l}$ as notation for the predicted group of top $l$ lines.
Note that, due to estimation and prediction errors,
the true rank-orders $\bm{r}^{o}_{l}$ of the lines in $T_{l}$
may not be $1,\ldots,l$.
We evaluate the accuracy by the average median of $\bm{r}^{o}_l$
over 1000 simulations. The smaller the average median is,
the better the predicted group is.
In the following, we study the accuracy of the first ten groups,
$T_{1},\ldots,T_{10}$, for different methods.

\begin{figure}[!h]
    \centering
    \includegraphics[width=0.8\textwidth]{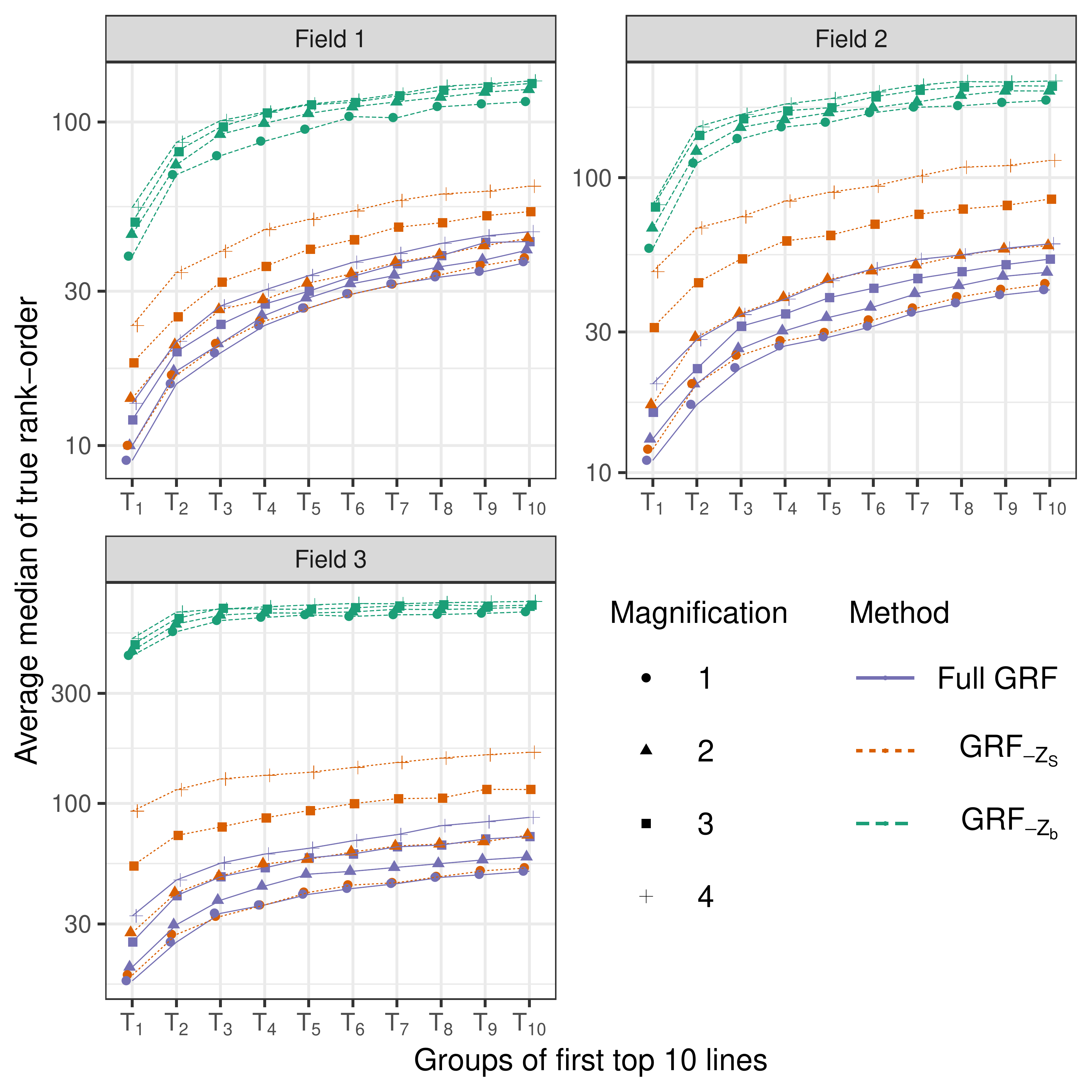}
    \caption{Rank-order analysis for \textsf{Data1} using three different methods under different magnification ($c$) of the spatial effects. The Y-axis is on a logarithmic scale.}
    \label{fig:rank}
\end{figure}

Figure \ref{fig:rank} shows the average median of $\bm{r}^{(o)}_{l}$ for $l=1,\ldots,10$ for the full GRF, $\text{GRF}_{-Z_{s}}$ and $\text{GRF}_{-Z_{b}}$ on each field. The horizontal axis represents different groups while the vertical axis represents the corresponding average median of $\bm{r}^{(o)}_{I}$.
We can see in Figure \ref{fig:rank} that the full GRF performs consistently better than $\text{GRF}_{-Z_{s}}$ and $\text{GRF}_{-Z_{b}}$ which suggests that accounting for either spatial or subpopulation effects improves selection of the best plant genotypes. 

\subsection{\textsf{Data2} Ranking}
For the wheat dataset \textsf{Data2}, we report the performances of $\text{GRF}_{-Z_{b}}$, $\text{GRF}_{-Z_{bs}}$ and CMLM based on $1000$ simulations, for each of the eight phenotypes.
We achieve similar conclusions as in \textsf{Data1}.
Due to space limitation, the details are relegated to
Section \ref{appsec:data2rank} of
the supplementary material.

\section{Discussion}\label{sec:discussion}
This paper investigates the problem of adjusting for spatial effects in genomic prediction. Our analysis of the maize dataset \textsf{Data1} and the wheat dataset \textsf{Data2} reveals the existence of spatial effects and heterogeneity across different subpopulation families. The spatial effects and heterogeneity, without proper treatment, can reduce the quality of phenotypic prediction and genotypic ranking. Under the Gaussian random field model, we propose an additive covariance matrix structure that incorporates genotype effects, spatial effects and subpopulation effects. We have also shown that by adjusting for spatial effects, we can improve the selection of top-performing plant genotypes.

As a guest Editor suggests, block cross-validation \citep{Roberts-Bahn-Ciuti17} could be a more practical cross-validation method when the data are spatially correlated because it produces cross-validation errors that better match the magnitude of errors expected in practice. However, its effect on the ranking of the models has not been well understood other than few recondite applications where the spatial domains are disconnected \citep{hao2020testing}. More exploration on this direction would be beneficial.

\section{Acknowledgment}
The authors thank the Joint Editor and the two reviewers whose constructive comments and suggestions led to a considerably improved version of the manuscript. The authors acknowledge the Iowa State University Plant Sciences Institute Scholars Program for financial support and the lab of Patrick S. Schnable and former graduate research assistant Sarah Hill--Skinner for collecting and sharing the maize data. This article is a product of the Iowa Agriculture and Home Economics Experiment Station, Ames, Iowa. Project No. IOW03617 is supported by USDA/NIFA and State of Iowa funds. Xiaojun Mao's research is partially supported by Shanghai Sailing Program 19YF1402800. Any opinions, findings, conclusions or recommendations expressed in this publication are those of the authors and do not necessarily reflect the views of the U.S. Department of Agriculture or the Science and Technology Commission of Shanghai Municipality.

\bibliographystyle{ECA_jasa}
\bibliography{GRF}
\newpage
\input{ASEGP_SM.tex}

\end{document}

%% file: ASEGP_SM.tex
\pagebreak
\begin{center}
\textbf{\Large Supplement to ``Adjusting for Spatial Effects in Genomic Prediction"}\\
{Xiaojun Mao, Somak Dutta, Raymond K. W. Wong and Dan Nettleton}
\end{center}
\setcounter{equation}{0}
\setcounter{figure}{0}
\setcounter{table}{0}
\setcounter{page}{1}
\setcounter{section}{0}
\makeatletter
\renewcommand{\thesection}{S\arabic{section}}
\renewcommand{\thesubsection}{\thesection.\arabic{subsection}}
\renewcommand{\theequation}{S\arabic{equation}}
\renewcommand{\thefigure}{S\arabic{figure}}
\renewcommand{\bibnumfmt}[1]{[S#1]}
\renewcommand{\citenumfont}[1]{S#1}
\renewcommand{\thetable}{S\arabic{table}}

\begin{center}
 \textbf{Abstract}
 \end{center}
This document provides supplementary material to the article ``Adjusting for Spatial Effects in Genomic Prediction" written by the same authors.
\vspace{0.5cm}


\section{Data pre-processing}\label{appsec:preprocess}

In this section, we provide a detailed description of the pre-processing procedures for the maize dataset \textsf{Data1} and the wheat dataset \textsf{Data2}.

\begin{figure}[!h]
	\centering
	\includegraphics[scale=0.6]{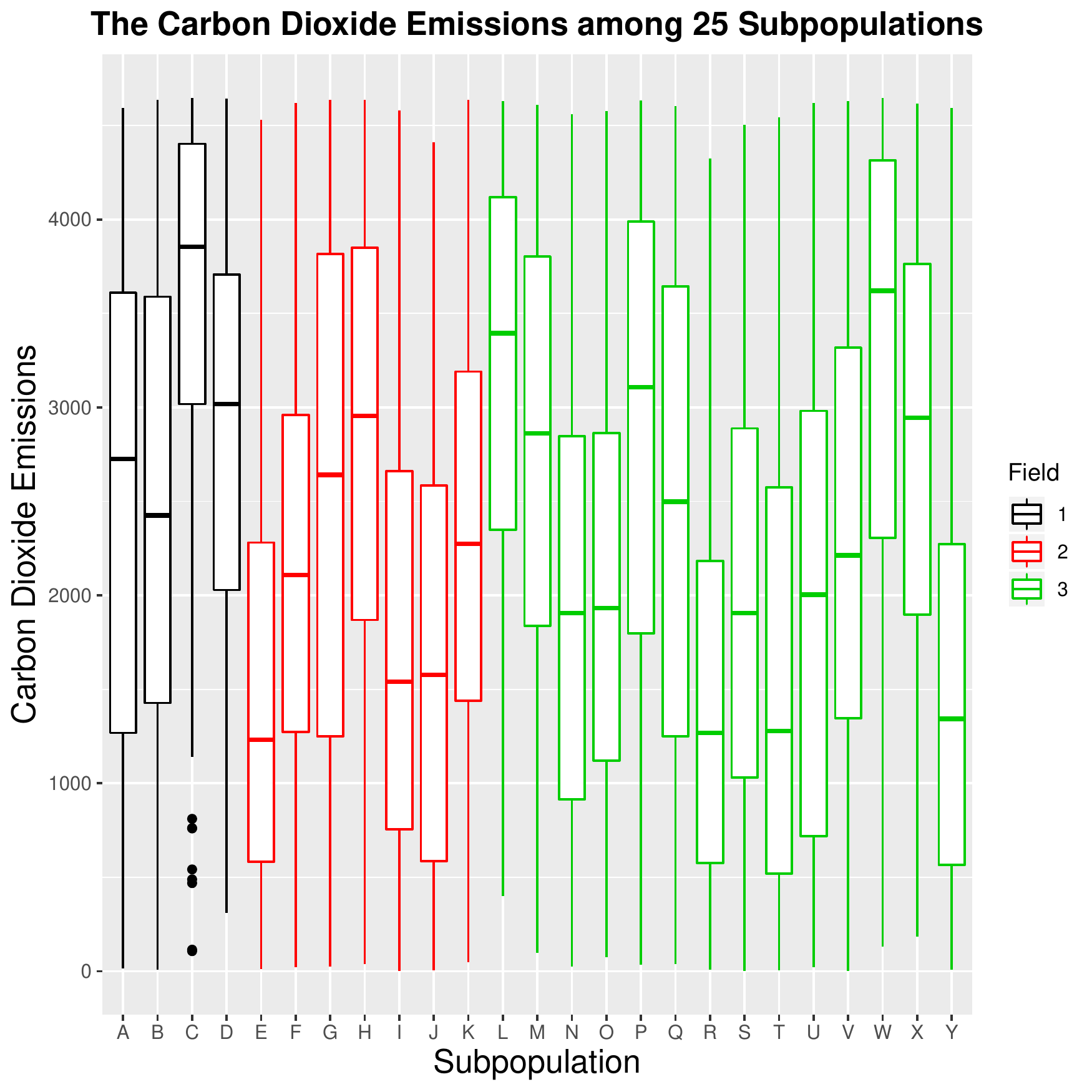}
	\caption{Carbon Dioxide Emissions among 25 subpopulations A through Y.}
	\label{appfig:carbonsub}
\end{figure}

Figure \ref{appfig:carbonsub} shows the boxplot of the carbon dioxide emissions of the 25 subpopulations.

For the maize dataset \textsf{Data1}, Table~\ref{apptab:miss} shows the average missing rate and average minor allele frequency (MAF) for SNP genotypes by chromosome after missing genotypes were imputed with LD-kNNi \citep{Money-Gardner-Migicovsky15}.
\begin{table}[!h]
	\caption{The average missing rate and average MAF after imputation for each chromosome.}
	\label{apptab:miss}
	\centering
	\footnotesize 
	\begin{tabular}{rrrrrrrrrrr}
		\hline
		\text{Chromosome} & 1 & 2 & 3  & 4 & 5 & 6 & 7 & 8  & 9 & 10\\
		\hline
		\text{Missing} & 0.1664 & 0.1607 & 0.1598 & 0.1657 & 0.1640 & 0.1592 & 0.1592 & 0.1642 & 0.1660 & 0.1592 \\
		\text{MAF} & 0.1186 & 0.1101 & 0.1074 & 0.1129 & 0.1177 & 0.1091 & 0.1118 & 0.1108 & 0.1153 & 0.1113 \\
		\hline
	\end{tabular}
\end{table}
After all missing SNP genotypes were imputed, we subdivided the 4660 observations into subsets corresponding to the three fields and, further, into subsets corresponding to the 25 subpopulations.

In each of these datasets, many SNP markers are identical to other SNP markers for all observations. We kept only one representative SNP marker in such cases and removed redundant SNPs. However, a huge number of SNP markers still remain. It is not only a computational burden for the marker kernel $\bm{C}_{g}$, but it also incorporates redundant information from highly correlated SNPs and useless information from SNPs unassociated with trait values. To get a more useful marker kernel, we used the R package $\mathrm{FarmCPU}$ \citep{Liu-Huang-Fan16} with default settings to select subsets of SNPs for computation of our marker kernel.  This selection of SNPs was done separately for each subpopulation and each field for \textsf{data1}, and separately for each trait for \textsf{data2}.  Table \ref{apptab:snpnumber} shows the total number of SNP markers before and after removing the duplicated markers, the number of selected SNP markers for \textsf{Data1} on Field 1, and the minimum MAF among the selected markers.

\begin{table}[!h]
	\caption{The total number of SNP markers before and after removing the duplicated markers, the number of selected SNP markers for \textsf{Data1} on Field 1, and the minimum MAF among the selected SNPs.}
	\label{apptab:snpnumber}
	\centering
	\footnotesize 
	\begin{tabular}{rrrrrrrrrrr}
		\hline
		\text{Chromosome} & 1 & 2 & 3  & 4 & 5 & 6 & 7 & 8  & 9 & 10  \\
		\hline
		\text{total} & 104827 & 81315 & 78369 & 73466 & 67423 & 58365 & 59577 & 59820 & 54013 & 50694 \\ 
		\text{unique} & 95065 & 71524 & 68790 & 64492 & 60645 & 51272 & 51803 & 52687 & 47238 & 44470 \\
		\text{selected} & 470 & 2666 & 789 & 300 & 503 & 290 & 580 & 343 & 124 & 401 \\
		\text{min MAF} & 0.0171 & 0.0237 & 0.0171 & 0.0197 & 0.0184 & 0.0276 & 0.0184 & 0.0237 & 0.0145 & 0.0184 \\
		\hline
	\end{tabular}
\end{table}
\noindent Table \ref{apptab:snpnumber} shows that, after selection, the number of SNP markers is greatly reduced. 


\section{Sensitivity to the parameter $\beta_{00}$}
\begin{table}[!h]
	\caption{The mean of accuracies for eight phenotypes by method $\text{GRF}_{-Z_{b}}$ ($\lambda_{00}=0.001$ and $\lambda_{00}=0.0001$) with full and selected SNPs based on $1000$ independent random partitions of the data into training ($86\%$) and test ($14\%$) sets.} \label{apptab:lambda}
	\centering
	\small
	\begin{tabular}{|c|c|c|c|c|c|}
		\cline{1-6}
		&  & \multicolumn{2}{|c|}{Full} & \multicolumn{2}{|c|}{Selected} \\
		\cline{1-6}
		\hline
		Water Supply & Phenotype & \multicolumn{2}{|c|}{$\text{GRF}_{-Z_{b}}$} & \multicolumn{2}{|c|}{$\text{GRF}_{-Z_{b}}$} \\ 
		\hline
		& $\lambda_{00}$ & $0.001$ & $0.0001$ & $0.001$ & $0.0001$ \\ 
		\hline
		FI & DH & 0.8667 & 0.8668 & 0.8160 & 0.8185 \\ 
		& GY & 0.8299 & 0.8305 & 0.8309 & 0.8314 \\ 
		& NKS & 0.7160 & 0.7133 & 0.7004 & 0.6976 \\ 
		& TKW & 0.8294 & 0.8284 & 0.7972 & 0.7968 \\ 
		MWS & DH & 0.8231 & 0.8249 & 0.7732 & 0.7729 \\ 
		& GY & 0.9275 & 0.9268 & 0.9269 & 0.9257 \\
		& NKS & 0.7248 & 0.7244 & 0.7143 & 0.7140 \\ 
		& TKW & 0.8867 & 0.8863 & 0.8696 & 0.8700 \\ 
		\hline
	\end{tabular}
\end{table}
\newpage
\section{\textsf{Data1} prediction results}
\begin{table}[!h]
	\caption{Average accuracies for \textsf{Data1} by five existing methods and two proposed methods ($\text{GRF}_{-Z_{bs}}$ and $\text{GRF}_{-Z_{b}}$) for each subpopulation based on $1000$ independent random partitions of the data into training ($80\%$) and test ($20\%$) sets. The highest average accuracy across methods for each subpopulation is shown in bold.} 
	\label{apptab:fieldsubresult}
	\centering
	\scriptsize 
	\begin{tabular}{|c|c|c|c|c|c|c|c|c|c|}
		\cline{1-10}
		\multicolumn{2}{|c|}{} & \multicolumn{7}{|c|}{Method} &  \\
		\cline{1-10}
		\hline
		Field & {\scriptsize Subpopulation} & CMLM & \multicolumn{2}{|c|}{RC} & \multicolumn{2}{|c|}{MVNG} & $\text{GRF}_{-Z_{bs}}$ & $\text{GRF}_{-Z_{b}}$ & $\widehat{\gamma}=\widehat{\sigma}_{s}^{2}/\widehat{\sigma}_{g}^{2}$ \\ 
		\hline
		& &  & RR & GAUSS & RR & GAUSS &  &  &  \\ 
		\hline
		1 & A & $\bm{0.5101}$ & 0.4568 & 0.4196 & 0.5098 & 0.5014 & 0.5036 & 0.5070 & 0.0220 \\ 
		& B & 0.4706 & 0.4077 & 0.3544 & $\bm{0.4715}$ & 0.4610 & 0.4657 & 0.4530 & 9e-04 \\ 
		& C & $\bm{0.5875}$ & 0.4594 & 0.4388 & 0.5855 & 0.5805 & 0.5788 & 0.5772 & 0.0000 \\ 
		& D & 0.4939 & 0.3139 & 0.0779 & 0.4933 & 0.4803 & 0.4886 & $\bm{0.5113}$ & 0.0249 \\ 
		\hline
		2 & E & 0.5300 & 0.2966 & 0.1407 & 0.5293 & 0.5295 & $\bm{0.5377}$ & 0.5308 & 0.0228 \\ 
		& F & 0.4939 & 0.4102 & 0.3782 & 0.4925 & 0.4847 & 0.4812 & $\bm{0.5564}$ & 0.0994 \\ 
		& G & 0.5314 & 0.4670 & 0.4424 & 0.5314 & 0.5278 & 0.5291 & $\bm{0.5466}$ & 0.0412 \\ 
		& H & 0.5250 & 0.4421 & 0.3324 & 0.5250 & 0.5281 & 0.5295 & $\bm{0.5527}$ & 0.0704 \\ 
		& I & 0.5694 & 0.4629 & 0.4235 & 0.5688 & 0.5689 & 0.5674 & $\bm{0.6097}$ & 0.0576 \\ 
		& J & 0.4947 & 0.4741 & 0.4588 & $\bm{0.4950}$ & 0.4916 & 0.4901 & 0.4841 & 0.0285 \\ 
		& K & $\bm{0.6179}$ & 0.4399 & 0.3275 & 0.6166 & 0.6044 & 0.6088 & 0.6014 & 0.0014 \\
		\hline
		3 & L & 0.5116 & 0.4354 & 0.2586 & 0.5104 & 0.5102 & 0.5102 & $\bm{0.5455}$ & 0.0498 \\ 
		& M & $\bm{0.5036}$ & 0.4427 & 0.4329 & 0.5024 & 0.5014 & 0.5005 & 0.5028 & 0.0276 \\ 
		& N & 0.5162 & 0.1193 & 0.0646 & 0.5148 & 0.5131 & 0.5084 & $\bm{0.5166}$ & 0.0375 \\ 
		& O & 0.5381 & 0.4741 & 0.2939 & 0.5373 & 0.5279 & 0.5338 & $\bm{0.6024}$ & 0.0670 \\ 
		& P & 0.4983 & 0.4857 & 0.4807 & 0.4966 & 0.5048 & 0.5153 & $\bm{0.6598}$ & 0.3922 \\ 
		& Q & 0.5562 & 0.4979 & 0.4632 & 0.5529 & 0.5508 & 0.5536 & $\bm{0.5959}$ & 0.1044 \\ 
		& R & 0.5563 & 0.4380 & 0.1916 & 0.5567 & 0.5563 & 0.5551 & $\bm{0.5802}$ & 0.0650 \\ 
		& S & 0.6193 & 0.6057 & 0.5845 & 0.6188 & 0.6166 & 0.6147 & $\bm{0.6482}$ & 0.0502 \\ 
		& T & 0.5892 & 0.4056 & 0.3383 & 0.5875 & 0.5837 & 0.5795 & $\bm{0.6143}$ & 0.0242 \\ 
		& U & 0.5886 & 0.4930 & 0.4807 & $\bm{0.5894}$ & 0.5795 & 0.5818  & 0.5815 & 0.0000 \\ 
		& V & 0.5160 & 0.3830 & 0.3097 & $\bm{0.5166}$ & 0.5076 & 0.5054 & 0.5162 & 0.0202 \\ 
		& W & 0.5110 & 0.4842 & 0.4325 & $\bm{0.5116}$ & 0.5049 & 0.5068 & 0.4959 & 0.0038 \\ 
		& X & 0.4990 & 0.4314 & 0.4078 & $\bm{0.4991}$ & 0.4960 & 0.4947 & 0.4861 & 0.0120 \\ 
		& Y & $\bm{0.5680}$ & 0.4432 & 0.3613 & 0.5662 & 0.5571 & 0.5593 & 0.5547 & 0.0015\\ 
		\hline
	\end{tabular}
\end{table}

\section{\textsf{Data2} prediction results}

\begin{table}[!h]
	\caption{Average accuracies for eight phenotypes by methods (CMLM, IB, RC(RR, GAUSS), MVNG(RR, GAUSS), SpATS, $\text{GRF}_{-Z_{bs}}$ and $\text{GRF}_{-Z_{b}}$) with full SNPs based on $1000$ independent random partitions of the data into training ($86\%$) and test ($14\%$) sets. The highest average accuracy across methods for each of the nine methods is printed in bold for each phenotype.}\label{apptab:wheatpred1}
	\centering
	\scriptsize 
	\begin{tabular}{|c|c|c|c|c|c|c|c|c|c|c|}
		\cline{1-11}
		&  & \multicolumn{9}{|c|}{Full} \\
		\cline{1-11}
		\hline
		WS & Phenotype & CMLM & IB & \multicolumn{2}{|c|}{RC} & \multicolumn{2}{|c|}{MVNG} & SpATS & $\text{GRF}_{-Z_{bs}}$ & $\text{GRF}_{-Z_{b}}$ \\ 
		\hline
		 &  &  &  & RR & GAUSS & RR & GAUSS & & & \\ 
		\hline
		FI & DH & $\bm{0.2091}$ & 0.0851 & 0.1863 & 0.1756 & 0.1084 & 0.1386 & 0.0672 & 0.1350 & 0.1435 \\ 
		& GY & -0.0346 & 0.6132 & 0.0662 & 0.0739 & 0.0934 & -0.0022 & 0.7410 & 0.0015 & $\bm{0.7575}$ \\ 
		& NKS & 0.0181 & 0.2355 & 0.2539 & $\bm{0.2776}$ & 0.1979 & 0.0999 & 0.1898 & 0.0638 & 0.2214 \\ 
		& TKW & 0.2720 & 0.3442 & 0.2682 & 0.2722 & 0.3217 & 0.3509 & 0.1252 & 0.3692 & $\bm{0.4040}$ \\ 
		MWS & DH & $\bm{0.2218}$ & 0.0945 & 0.1587 & 0.1571 & 0.1017 & 0.1458 & 0.1157 & 0.1516 & 0.2174 \\ 
		& GY & -0.0264 & 0.7734 & 0.0146 & 0.0135 & -0.0273 & -0.0110 & 0.8946 & 0.0183 & $\bm{0.8988}$ \\ 
		& NKS & -0.0033 & 0.2149 & 0.2487 & 0.2576 & 0.1328 & 0.1005 & 0.3015 & 0.1095 & $\bm{0.3112}$ \\ 
		& TKW & 0.2354 & 0.5530 & 0.2350 & 0.2394 & 0.2710 & 0.2869 & 0.5938 & 0.3132 & $\bm{0.6443}$ \\ 
		\hline
	\end{tabular}
\end{table}

\begin{table}[!h]
	\caption{Average accuracies for eight phenotypes by methods (CMLM, IB, RC(RR, GAUSS), MVNG(RR, GAUSS), SpATS, $\text{GRF}_{-Z_{bs}}$ and $\text{GRF}_{-Z_{b}}$) with selected SNPs based on $1000$ independent random partitions of the data into training ($86\%$) and test ($14\%$) sets. The highest average accuracy across methods for each of the nine methods is printed in bold for each phenotype.}\label{apptab:wheatpred2}
	\centering
	\scriptsize 
	\begin{tabular}{|c|c|c|c|c|c|c|c|c|c|c|}
		\cline{1-11}
		&  & \multicolumn{9}{|c|}{Selected} \\
		\cline{1-11}
		\hline
		WS & Phenotype & CMLM & IB & \multicolumn{2}{|c|}{RC} & \multicolumn{2}{|c|}{MVNG} & SpATS & $\text{GRF}_{-Z_{bs}}$ & $\text{GRF}_{-Z_{b}}$ \\ 
		\hline
		 &  &  &  & RR & GAUSS & RR & GAUSS & & & \\ 
		\hline
		FI & DH & 0.1637 & 0.0772 & 0.1540 & 0.1867 & $\bm{0.1928}$ & 0.1341 & 0.0672 & -0.0906 & -0.0234 \\ 
		& GY & -0.0396 & 0.6184 & 0.1395 & 0.1350 & 0.1505 & -0.0112 & 0.7410 & -0.1080 & $\bm{0.7618}$ \\ 
		& NKS & 0.0961 & 0.2070 & 0.1922 & $\bm{0.2588}$ & 0.1510 & 0.0958 & 0.1898 & -0.0095 & 0.1881 \\ 
		& TKW & 0.1997 & 0.2030 & 0.1096 & 0.2542 & 0.2666 & $\bm{0.2999}$ & 0.1252 & 0.0008 & 0.1442 \\ 
		MWS & DH & 0.1443 & 0.0317 & 0.0891 & $\bm{0.1521}$ & 0.0604 & 0.1406 & 0.1157 & -0.0198 & 0.1236 \\ 
		& GY & -0.0324 & 0.7751 & 0.1068 & 0.1048 & 0.0338 & -0.0109 & 0.8946 & -0.0843 & $\bm{0.8979}$ \\ 
		& NKS & -0.0322 & 0.2342 & 0.2433 & 0.2626 & 0.1073 & 0.1219 & $\bm{0.3015}$ & 0.0031 & 0.2925 \\ 
		& TKW & 0.2546 & 0.5276 & 0.2275 & 0.2242 & 0.2471 & 0.2635 & $\bm{0.5938}$ & -0.0002 & 0.5870 \\ 
		\hline
	\end{tabular}
\end{table}

\begin{table}[!h]
	\caption{The spatial parameter estimates for eight phenotypes by method $\text{GRF}_{-Z_{b}}$ with full and selected SNPs.} \label{apptab:wheatspatial}
	\centering
	\footnotesize
	\begin{tabular}{|c|c|c|c|c|c|c|c|}
		\cline{1-8}
		&  & \multicolumn{3}{|c|}{Full} & \multicolumn{3}{|c|}{Selected} \\
		\cline{1-8}
		\hline
		Water Supply & Phenotype & $\beta_{01}$ & $\beta_{10}$ & $\widehat{\gamma}$ & $\beta_{01}$ & $\beta_{10}$ & $\widehat{\gamma}$ \\ 
		\hline
		FI & DH & 0.0344 & 0.4656 & 0.0277 & 0.0142 & 0.4858 & 0.0165 \\ 
		& GY & 0.0587 & 0.4413 & 2.4231 & 0.0612 & 0.4388 & 2.6715\\ 
		& NKS & 0.0077 & 0.4923 & 0.1670 & 0.0116 & 0.4884 & 0.1629 \\ 
		& TKW & 0.0264 & 0.4736 & 0.0877 & 0.0270 & 0.4730 & 0.0724 \\ 
		MWS & DH & 0.0394 & 0.4606 & 0.0451 & 0.0333 & 0.4667 & 0.0597 \\ 
		& GY & 0.0644 & 0.4356 & 4.5171 & 0.0688 & 0.4312 & 4.7537 \\
		& NKS & 0.0596 & 0.4404 & 0.2502 & 0.0569 & 0.4431 & 0.2510 \\ 
		& TKW & 0.0861 & 0.4139 & 0.4125 & 0.1109 & 0.3891 & 0.4787 \\ 
		\hline
	\end{tabular}
\end{table}

\newpage
\section{\textsf{Data2} ranking}
\label{appsec:data2rank}

\begin{figure}[!h]
	\centering
	\includegraphics[scale=0.4]{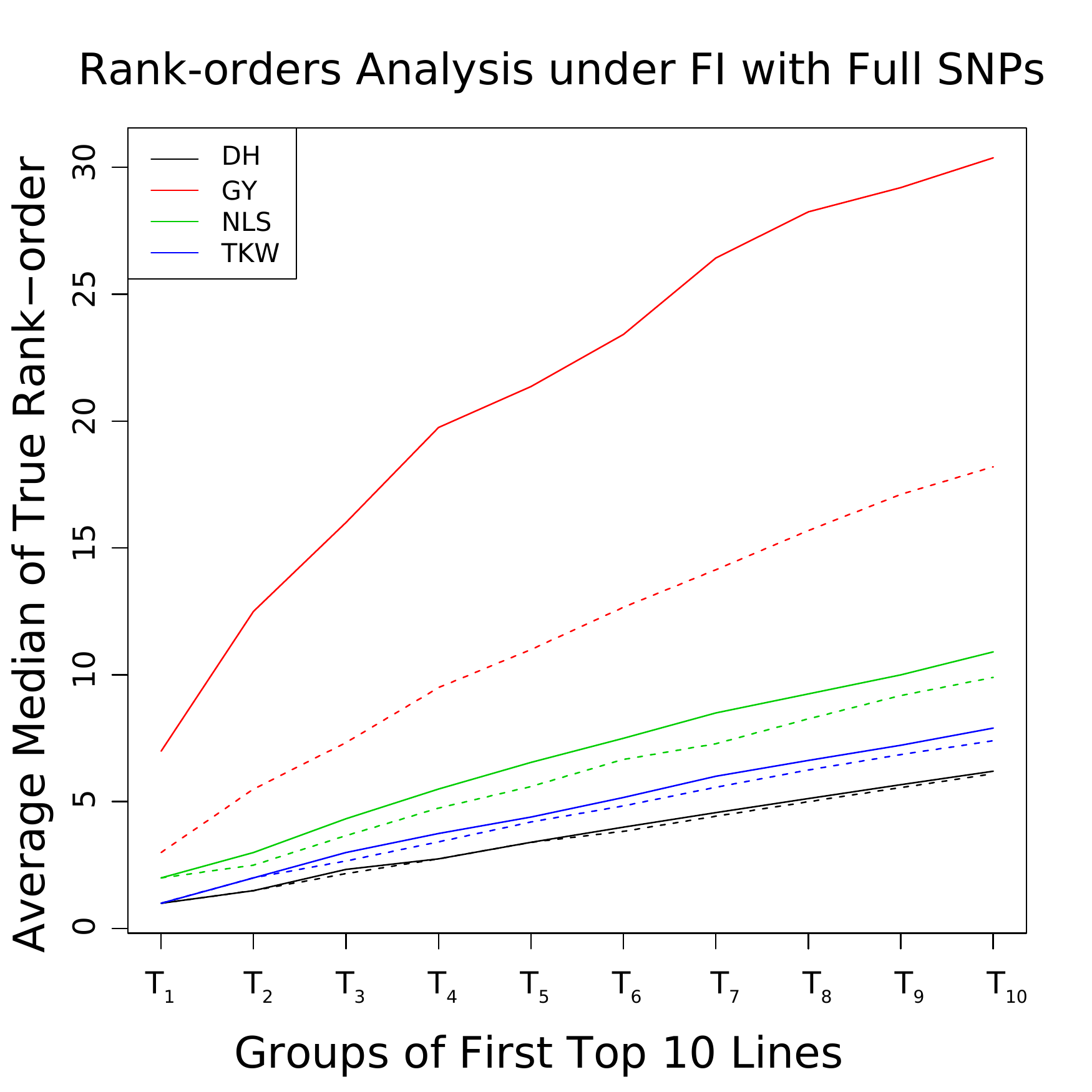}
	\includegraphics[scale=0.4]{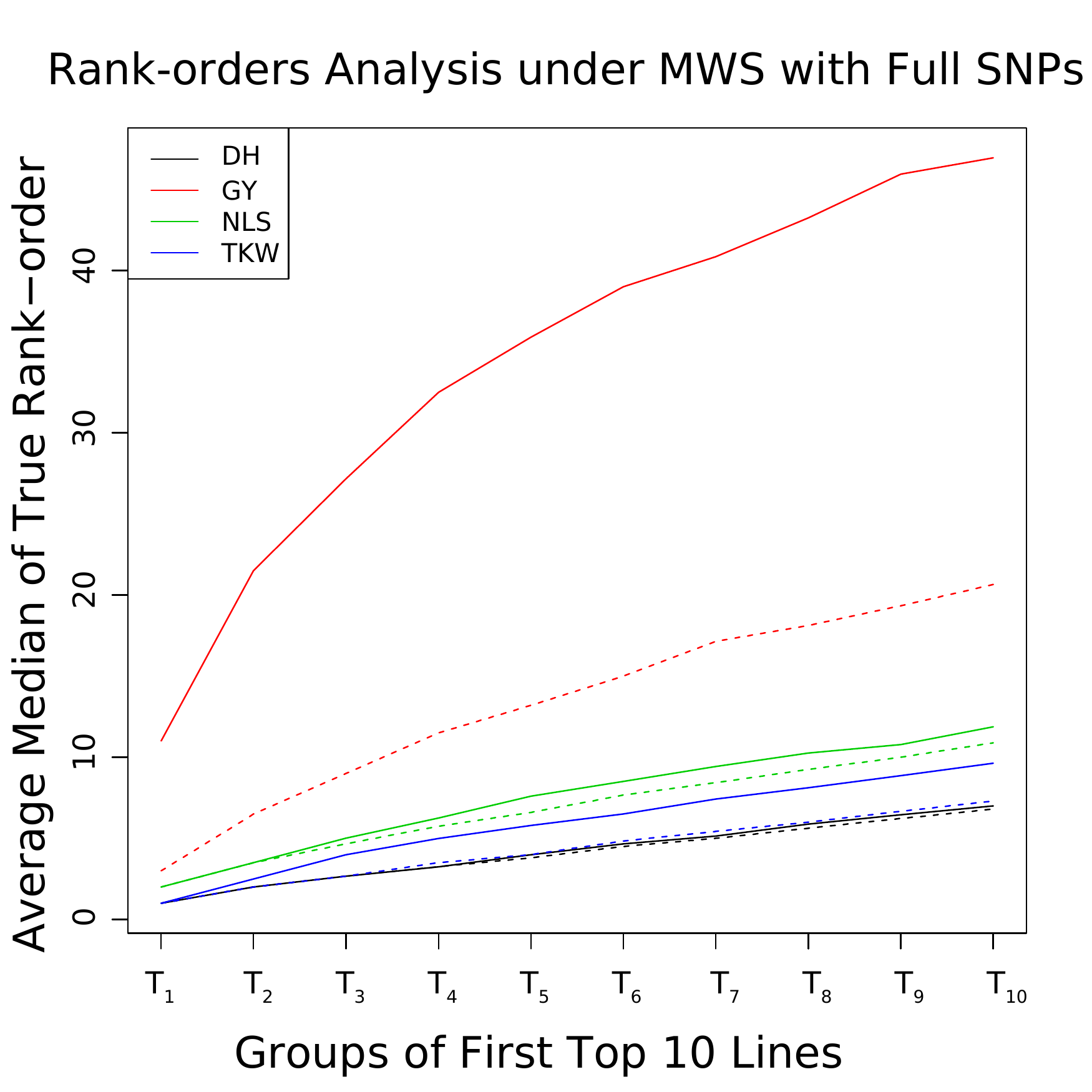}
	\includegraphics[scale=0.4]{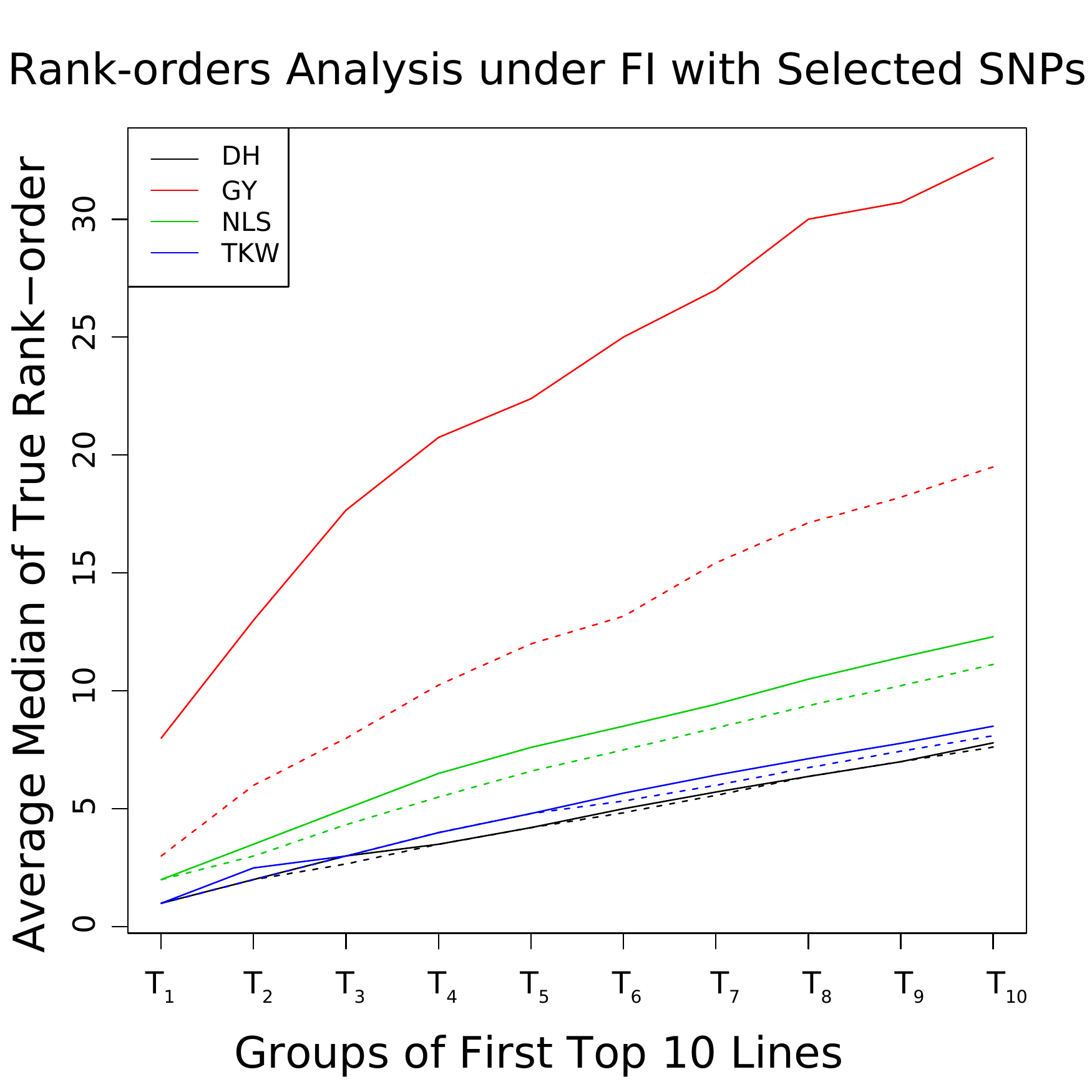}
	\includegraphics[scale=0.4]{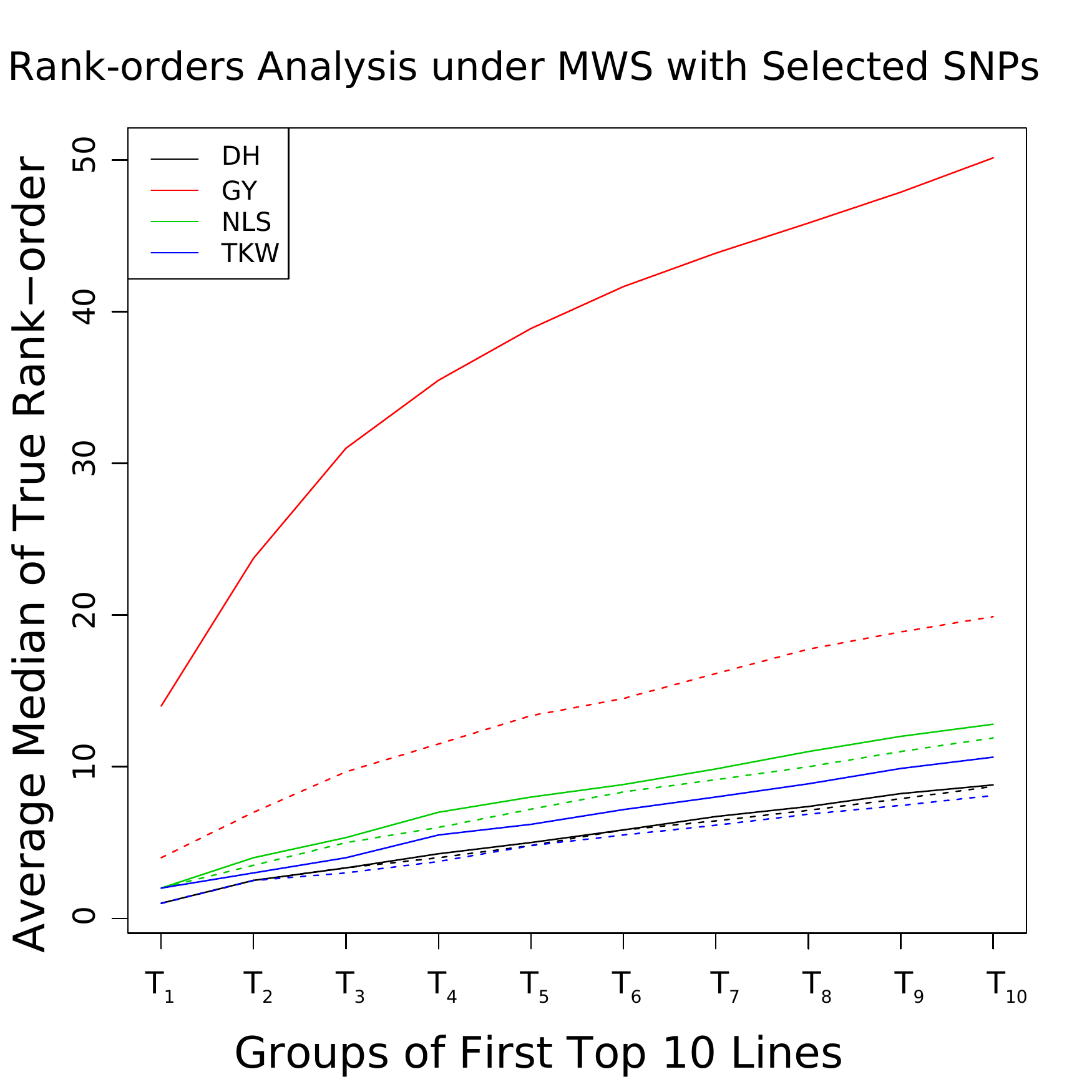}
	\caption{Comparisons of $\text{GRF}_{-Z_{bs}}$ and $\text{GRF}_{-Z_{b}}$ with full and selected SNPs under full irrigated (FI) conditions and mild water stress. The solid lines are for $\text{GRF}_{-Z_{bs}}$, while dashed lines are for $\text{GRF}_{-Z_{b}}$.}
	\label{appfig:rankwheat}
\end{figure}

Table \ref{apptab:rankorderwheat} reports both the prediction accuracies and Spearman's rank-order correlations. As before, these two measurements are highly correlated. The results show that with strong spatial effects, i.e., phenotype GY under full irrigated (FI) conditions and mild water stress, $\text{GRF}_{-Z_{b}}$ is much better than $\text{GRF}_{-Z_{bs}}$ in terms of prediction accuracies and the similarities of rank-orders with the true rank-order $\bm{r}^{(o)}$. We can see in Figure \ref{appfig:rankwheat} that $\text{GRF}_{-Z_{b}}$ is consistently better than $\text{GRF}_{-Z_{bs}}$. This provides the evidence that accounting for spatial effects improves selection of the best plant genotypes.

\begin{table}[!h]
	\caption{Average prediction accuracies and Spearman's rank-order correlations ($\rho_s$) based on 1000 simulations for different phenotypes by $\text{GRF}_{-Z_{bs}}$ and $\text{GRF}_{-Z_{b}}$ with full and selected SNPs under full irrigated (FI) conditions and mild water stress.}
	\centering
	\scriptsize
	\label{apptab:rankorderwheat}
	\begin{tabular}{|c|c|c|c|c|c|c|c|c|c|}
		\cline{1-10}
		&  & \multicolumn{4}{|c|}{Full} & \multicolumn{4}{|c|}{Selected} \\
		\hline
		& & \multicolumn{2}{|c|}{Accuracies} & \multicolumn{2}{|c|}{$\rho_s$}
		& \multicolumn{2}{|c|}{Accuracies} & \multicolumn{2}{|c|}{$\rho_s$} \\
		\cline{1-10}
		\hline
		WS & Phenotype & $\text{GRF}_{-Z_{bs}}$ & $\text{GRF}_{-Z_{b}}$ & $\text{GRF}_{-Z_{bs}}$ & $\text{GRF}_{-Z_{b}}$ & $\text{GRF}_{-Z_{bs}}$ & $\text{GRF}_{-Z_{b}}$ & $\text{GRF}_{-Z_{bs}}$ & $\text{GRF}_{-Z_{b}}$\\ 
		\hline
		FI & DH & 0.9870 & 0.9895 & 0.9851 & 0.9879 & 0.9604 & 0.9606 & 0.9558 & 0.9562 \\ 
		& GY & 0.7241 & 0.8401 & 0.7056 & 0.8261 & 0.7123 & 0.8334 & 0.6940 & 0.8196 \\ 
		& NKS & 0.9148 & 0.9302 & 0.9056 & 0.9222 & 0.9067 & 0.9184 & 0.8972 & 0.9098 \\
		& TKW & 0.9622 & 0.9697 & 0.9574 & 0.9658 & 0.9505 & 0.9558 & 0.9446 & 0.9504 \\ 
		\hline
		MWS & DH & 0.9749 & 0.9783 & 0.9714 & 0.9751 & 0.9467 & 0.9503 & 0.9405 & 0.9443 \\ 
		& GY & 0.6257 & 0.8201 & 0.6060 & 0.8052 & 0.6251 & 0.8195 & 0.6077 & 0.8051 \\ 
		& NKS & 0.9036 & 0.9184 & 0.8938 & 0.9099 & 0.8927 & 0.9064 & 0.8818 & 0.8966 \\ 
		& TKW &0.9336 & 0.9702 & 0.9258 & 0.9660 & 0.9210 & 0.9533 & 0.9122 & 0.9478 \\ 
		\hline 
	\end{tabular}
\end{table}

